\documentclass[sn-basic]{sn-jnl}% Basic Springer Nature Reference Style/Chemistry Reference Style
\usepackage{graphicx}%
\usepackage{multirow}%
\usepackage{amsmath,amssymb,amsfonts}%
\usepackage{amsthm}%
\usepackage{mathrsfs}%
\usepackage[title]{appendix}%
\usepackage{xcolor}%
\usepackage{textcomp}%
\usepackage{manyfoot}%
\usepackage{booktabs}%
\usepackage{algorithm}%
\usepackage{algorithmicx}%
\usepackage{algpseudocode}%
\usepackage{listings}%
\usepackage{natbib}
\usepackage{booktabs}
\usepackage{bm}

\def\be{\begin{equation}}
\def\ee{\end{equation}}
\def\bea{\begin{eqnarray}}
\def\eea{\end{eqnarray}}
\setcitestyle{aysep={},yysep={;}}

\theoremstyle{thmstyleone}%
\theoremstyle{thmstyletwo}%

\theoremstyle{thmstylethree}%

\begin{document}

\title[GRB Constraints Using Machine Learning]
{Model-independent Gamma-Ray Bursts Constraints on Cosmological Models Using  Machine Learning }

%%=============================================================%%
%% Prefix	-> \pfx{Dr}
%% GivenName	-> \fnm{Joergen W.}
%% Particle	-> \spfx{van der} -> surname prefix
%% FamilyName	-> \sur{Ploeg}
%% Suffix	-> \sfx{IV}
%% NatureName	-> \tanm{Poet Laureate} -> Title after name
%% Degrees	-> \dgr{MSc, PhD}
%% \author*[1,2]{\pfx{Dr} \fnm{Joergen W.} \spfx{van der} \sur{Ploeg} \sfx{IV} \tanm{Poet Laureate}
%%                 \dgr{MSc, PhD}}\email{iauthor@gmail.com}
%%=============================================================%%

\author[1,2]{\fnm{Bin} \sur{Zhang}}\email{binzhang@gznu.edu.cn}

\author[1]{\fnm{Huifeng} \sur{Wang}}\email{}
%\equalcont{These authors contributed equally to this work.}

\author[1]{\fnm{Xiaodong} \sur{Nong}}\email{}
%\equalcont{These authors contributed equally to this work.}
\author[1]{\fnm{GuangZhen} \sur{Wang}}\email{}

\author[3,4]{\fnm{Puxun} \sur{Wu}}\email{pxwu@hunnu.edu.cn}

\author*[1]{\fnm{Nan} \sur{Liang}}\email{liangn@bnu.edu.cn}

\affil*[1]{\orgdiv{Key Laboratory of Information and Computing Science Guizhou Province (School of Cyber Science and Technology)}, \orgname{Guizhou Normal University}, \orgaddress{\city{Guiyang}, \postcode{Guizhou 550025}, \country{China}}}

\affil*[2]{\orgdiv{School of Mathematical Sciences}, \orgname{Guizhou Normal University}, \orgaddress{\city{Guiyang}, \postcode{Guizhou 550025}, \country{China}}}

%\affil[3]{\orgdiv{ School of Cyber Science and Technology}, \orgname{Guizhou Normal University}, \orgaddress{\city{Guiyang}, \postcode{Guizhou 550025}, \country{China}}}

%\affil[3]{\orgdiv{Joint Center for FAST Sciences Guizhou Normal University Node}, \orgaddress{\city{Guiyang}, \postcode{Guizhou 550025}, \country{China}}}
\affil*[3]{\orgdiv{Department of Physics and Synergistic Innovation Center for Quantum Effects and Applications}, \orgname{Hunan Normal University}, \orgaddress{\city{Changsha}, \postcode{Hunan 410081}, \country{China}}}
\affil[4]{\orgdiv{Institute of Interdisciplinary Studies}, \orgname{Hunan Normal University}, \orgaddress{\city{Changsha}, \postcode{Hunan 410081}, \country{China}}}

%%==================================%%
%% sample for unstructured abstract %%
%%==================================%%

\abstract{In this paper, we calibrate  the luminosity relation of  gamma-ray bursts (GRBs)  with the  machine learning (ML) algorithms %for reconstructing distance-redshift relation
from the Pantheon+  sample of type Ia supernovae in a cosmology-independent way. By using K-Nearest Neighbors (KNN) and Random Forest (RF) selected with the best performance in the ML algorithms, we calibrate the  Amati relation (\unboldmath{$E_{\rm p}$-$E_{\rm iso}$}) relation with the  A219 sample %at low redshift
to construct the Hubble diagram of GRBs.
Via the Markov Chain Monte Carlo numerical method with GRBs at high redshift  and latest observational Hubble data, we %obtain \unboldmath{$\Omega_{\rm m}$ = $0.312^{+0.088}_{-0.068}$, $h$ = $0.653^{+0.052}_{-0.073}$, $w_0$ = $-0.99^{+0.75}_{-0.43}$, $w_a$ = $-0.97^{+0.58}_{-0.58}$} at 1-\unboldmath{$\sigma$} confidence level for the Chevallier-Polarski-Linder model in a flat space, which
find the  results of constraints on cosmological models by using KNN and RF algorithms are consistent with those obtained from GRBs calibrated by using the Gaussian Process.
}


%%\pacs[JEL Classification]{D8, H51}

%%\pacs[MSC Classification]{35A01, 65L10, 65L12, 65L20, 65L70}

\maketitle

\section{Introduction}
%In practice, astrophysical objects often struggle to reach the state of a standard candle due to various disturbances and precision problems with observation instruments.
Type Ia supernovae (SNe Ia) have been often used as a standard candle with the maximum redshift observed at $z\sim2.3$ \citep{Scolnic2018,Scolnic2022}. Therefore, observations of luminous objects at higher redshift than SNe Ia are required to explore the cosmic evolution at the high-redshift region.
Gamma-ray bursts (GRBs) are the most intense bursts of high-energy gamma rays in a short time at high redshifts (the maximum redshift of GRB can reach at $z\sim9$ \citep{Cucchiara2011}). % which are proposed as supplementary tools to research the expansion of the Universe.
%At present, the maximum redshift of GRB can reach at $z\sim9$ \cite{Cucchiara2011}, while the maximum redshift observed for SNe Ia is about $z\sim2$ \cite{Scolnic2022}.  Therefore, GRBs can be used to probe the universe at high-redshift beyond SNe Ia.
%The so-called Amati relation, which connects the spectral peak energy and the isotropic equivalent radiated energy (the $E_{\rm p}$-${E}_{\rm iso}$ correlation) of GRBs, has been proposed by \cite{Amati2002}.
Utilizing the  GRB's luminosity relations,  which are connections between measurable properties of the instantaneous %gamma-ray
emission and the luminosity or energy
\citep{Fenimore2000,Norris2000,Ghirlanda2004a,Yonetoku2004,Liang2005,Firmani2006,Dainotti2008,Tsutsui2009a},
GRBs have been used as cosmic probe
to study the evolutionary history of our universe and the properties of dark energy\citep{Schaefer2003,Dai2004,Ghirlanda2004b,Firmani2005,Xu2005,Liang2006,Wang2006,Ghirlanda2006,Schaefer2007}.
%See \cite{Moresco2022} for reviews.
%\footnote{For recent of GRB luminosity relations and the applications in cosmology, see e.g.
%\citep{Izzo2015,Dainotti2018,Dainotti2020,Dainotti2022a,Xu2021,Xu2022,Xu2023,Hu2021,Wang2022,Liu2022a,Li2023}, and \citep{Shirokov2020,Khadka2020,Khadka2021,Luongo2020,Demianski2021,Cao2022a,Cao2022b,Liu2022b,Liang2022,Dainotti2022b,Dainotti2022c,Dainotti2023,LZL2023,Kumar2023,Mu2023,Bargiacchi2023}.
%}

In the early studies of GRB cosmology \citep{Dai2004,Schaefer2007}, the luminosity relations of GRBs  had usually been calibrated  by assuming  a certain cosmological model. %due to the lack of a low-redshift sample.
Thus, the so-called circularity problem is encountered \citep{Ghirlanda2006}.
In order to avoid this circularity problem, \cite{Liang2008} proposed a cosmological model-independent method to calibrate GRBs at low redshift  interpolated from SNe Ia and built the GRB Hubble diagram at high redshift. Following the interpolation method used in \citep{Liang2008},
many works have constrained  cosmological models with GRBs without any cosmological assumption, see, i. e. \citep{Capozziello2008,Capozziello2009,Wei2009,Wei2010,Liang2010,Liang2011,Wang2016,Liu2022b}.
On the other hand, the simultaneous method \citep{Amati2008,Wang2008} in which the parameters of the relationship and the cosmological model fitting simultaneously has been proposed to avoid the circularity problem.
%However, a particular cosmological model is still required in doing the joint fitting.
Recently, it is found that the GRB relation parameters are almost identical in all cosmological models, %fitted by the simultaneous method
which seems to indicate that GRBs can be standardized within error bars \citep{Khadka2020}.
It should be notice that GRB luminosity relations can be calibrated by using other observations. For example,
\cite{Amati2019} proposed  to calibrate GRB correlations by using the observational Hubble data (OHD) obtained with the cosmic chronometers (CC) method  fitted by the B\'ezier parametric, and built up a Hubble diagram consisting of 193 GRBs with the Amati relation %which connects the spectral peak energy and the isotropic equivalent radiated energy
(the $E_{\rm p}$-${E}_{\rm iso}$ correlation)\citep{Amati2002}.\footnote{Besides the calibration method by using SN Ia and OHD, the mock data of gravitational waves (GWs) \citep{Wang2019}, quasar sample \citep{Dai2021} and the angular diameter distances of galaxy clusters \citep{Gowri2022} have also been used  to calibrate GRBs. %to find that no evolution of redshift with the Amati relation. \cite{Gowri2022} used the angular diameter distances of galaxy clusters to calibrated the Amati relation at $z < 0.9$.
}
Following this method,
several works have constrained  cosmological models with the Amati relation calibrated by OHD \citep{Montiel2021,Luongo2021b,Luongo2023,Muccino2023}.
\footnote{For recent GRB luminosity relations and the applications in cosmology, see e.g.
\cite{Izzo2015,Dainotti2018,Dainotti2020,Dainotti2022a,Xu2021,Xu2022,Xu2023,Hu2021,Wang2022,Liu2022a,Li2023}, and \cite{Shirokov2020,Khadka2020,Khadka2021,Luongo2020,Demianski2021,Cao2022a,Cao2022b,Liu2022b,Liang2022,Dainotti2022b,Dainotti2022c,Dainotti2023,LZL2023,Kumar2023,Mu2023,Bargiacchi2023,Dinda2023},
and \cite{Shah2024, Cao2024, WLL2024}. %\cite{}.
For reviews, see \cite{Luongo2021a,Moresco2022}.}

The reconstruction from cosmological data in the calibration of GRBs  can be constructed in several ways.
Similar to the interpolation method used in \citep{Liang2008} and the B\'ezier parametric used in \citep{Amati2019}, GRBs are calibrated from the local data by using the polynomial fitting \citep{Kodama2008,Tsutsui2009b}, an iterative procedure \citep{LiangZhang2008}, the local regression \citep{Cardone2009,Demianski2017a}, the cosmography methods \citep{Capozziello2010,Gao2012}, a two-steps method minimizing the use of SNe Ia \citep{Izzo2015,Muccino2021}, and the Pad\'e approximation method \citep{Liu2015}.
%The luminosity relations of GRBs can be calibrated with the local data by the similar methods, see e.g.  \citep{Capozziello2008,Kodama2008,Tsutsui2009b,Cardone2009,Capozziello2010,Liang2010,Liang2011,Wei2010,WangDai2011,Gao2012,Liu2015,Wang2016,Demianski2017b}.
Recently, the non-parametric method has been addressed to reconstruction of the dark
energy, which can effectively reduce the errors of reconstructed results compared to the approaches mentioned in the above.  \cite{Postnikov2014} studied  the evolution of the cosmological
equation of state in a nonparametric way %, without imposing any a priori functional form
with %SNe Ia, BAO and
high redshift GRBs.

Gaussian Process (GP) is a powerful nonlinear interpolating tool without the need of specific models or parameters, which is a fully Bayesian approach that describes a distribution over functions with a generalization of Gaussian distributions to function space \citep{Seikel2012a}. GP approach has been used in various cosmological studies, see, e.g. \citep{Seikel2012a,Seikel2012b,Seikel2013,Busti2014,Li2018}. %In these papers, functions of the Hubble parameter with respect to the redshift and the distance-redshift relation are frequently reconstructed from expansion rate measurements and Type Ia supernovae (SNe Ia), respectively.
%It is important to note that GP has certain limitation.
However, in GP analysis, it is typically assumed that the errors in observational data follow a Gaussian distribution \citep{Seikel2012a}, which may pose a substantial limitation when reconstructing functions from data.
\cite{Wei2017} found that GP exhibit sensitivity to the fiducial Hubble constant $H_0$ and the results are significantly impacted by $H_0$.
\cite{Zhou2019} proposed that GP should be used with caution for the reconstruction of OHD and SNe Ia.
Furthermore, the results can be affected by the choose of the kernel functions, and there are a lot of kernel functions available that we can choose. %This sensitivity raises concerns about the reliability of GP in reconstructing $H(z)$.
Machine Learning (ML) algorithms are a set of technologies that learn to make predictions and decisions by training with a large amount of the observational data, which are a collection of processing units designed to identify underlying relationships in input data; therefore, when an appropriate network is chosen, the model created using ML can accurately depict the distribution of the input data in a completely data-driven way.
The ML methods have shown outstanding performance in solving cosmological problems in both accuracy and efficiency to provide powerful tools and methods for cosmological research \citep{Fluri2018,Fluri2019,Arjona2020,Wang2020,Wang2021,Xu2022,Escamilla-Rivera2020,Luongo2021b,Bengaly2023}.
%Through machine learning techniques, researchers can explore more deeply important questions.
Genetic Algorithms (GA) has been used to investigate the redshift evolution of the Cosmic Microwave Background (CMB) temperature \citep{Arjona2020}, the distance duality relations (DDR) with gravitational wave (GW) \citep{Hogg2020,Arjona2021},
and the late-time cosmological tensions using redshift-space distortion data in the low-redshift background to show that phantom dark energy is more preferable than the cosmological constant \citep{Gangopadhyay2023}.
\cite{Wang2020} proposed a new nonparametric approach for reconstructing a function from observational data using an Artificial Neural Network (ANN). % which has no assumptions about the data and is a completely data-driven approach.
\cite{Escamilla-Rivera2020} used the Recurrent Neural Networks (RNN) and the Bayesian Neural Networks (BNN) methods to reduce the computation load of expensive codes for dark energy models; these methods have subsequently been used to calibrate the GRB relations \citep{Escamilla-Rivera2022,Tang2021,Tang2022}.

Recently,
\cite{Luongo2021b} explored three machine learning treatments (linear regression, neural network, and random forest) based on B\'ezier polynomials to alleviate the circularity problem with the Amati relation.
The main issue in the calibration of GRBs is that we do not know a priorithe the correct curve to fitting data.
The overall advantage on using machine learning has been discussed in \citep{Luongo2021b}:
i) Healing degeneracy and over-fitting issues.
Multiple models can fit the same data will lead to degeneracy in fitting data approaches, and the overall approach of ML overcomes those issues due to interpolation, polynomials with generic over-fitting treatments.
ii) Speeding up the process of data adaption.
ML can maintain the consistency of data, which automatically encapsulates data without postulation over the shapes and orders.
%ML is a consistency of the data, which automatically encapsulates data without human postulation over their shapes and orders.
The complexity of ML models turns out to intimately related to the number of data points.
Therefore, the overall process of calibration can be improved.
\cite{Bengaly2023} deployed ML algorithms to measure the $H_0$ through regression analysis as Extra-Trees, ANN, Gradient Boosting, and Support Vector Machines (SVM), and found that the SVM exhibits the best performance in terms of bias-variance tradeoff in most cases, showing itself a competitive cross-check to GP.

More recently,
\cite{Khadka2021} compiled  a total 220 GRBs (the A220 sample) %in which a data set of 118 GRBs (the A118 sample) with the smallest intrinsic dispersion
to  derive the correlation and cosmological model parameter constraints simultaneously.
By using the GP method,
\cite{Liang2022}  calibrated the Amati relation with the A219 GRB sample\textbf{\footnote{Removed GRB051109A, which are counted twice in the A220 sample \citep{Khadka2021}.} } from the Pantheon sample \citep{Scolnic2018} which contains 1048 SNe, and constrained cosmological models in flat space with GRBs at high redshift and OHD via the Markov Chain Monte Carlo (MCMC) numerical method. % obtained by the cosmic chronometers (CC) method. %, which related the evolution of differential ages of passive galaxies at different redshifts \citep{Jimenez2002}.
\cite{LZL2023} calibrated GRBs from the latest OHD to construct the GRB Hubble diagram and constrained Dark Energy models with GRBs at high redshift and SNe Ia in a flat space.
\cite{Mu2023} used the Pantheon+ sample \citep{Scolnic2022}, which contains 1701 SNe light curves of 1550 spectroscopically confirmed SNe Ia, for calibrating the Amati relation to reconstruct cosmography parameters.
\cite{Xie2023} used the Pantheon+ sample to calibrate the Amati relation from the latest 221 GRB sample \citep{Jia2022}. % including 49 GRBs from Fermi catalog.

In this work, we calibrate the Amati relation with the A219 GRB data \citep{Liang2022} at low redshift from the Pantheon+ SNe Ia sample \citep{Scolnic2022} using the ML algorithms.
%Linear Regression, Lasso Regression, Bayesian Ridge, Decision Trees, Random Forest,
%Support Vector Regression, K-Nearest Neighbors, Gradient Boosting Decision Tree, and Multi-layer Perceptron, etc.)
%By the ANN+BNN architectures,
%We obtain the GRB Hubble diagram at high redshift and .
%\textbf{We also use a join network with ANN and BNN to consider the covariance errors of SN Ia and the uncertainty of the reconstruction.}
Combining the high redshift GRB data with the latest OHD, we constrain cosmological models % ( the $\Lambda$CDM model, the $w$CDM model)
in a flat space with MCMC numerical method. We also compare the results of ML and GP methods. %combine two likelihood methods \citep{D'Agostini2005,Reichart2001} to constrain cosmological models.

\section{Reconstructing the apparent magnitude redshift Relation from Pantheon+ \label{section2}}

In this section, we use ML algorithms to fit SNe Ia data set to reconstruct the apparent magnitude-redshift ($m-z$) relation. %to calibrate GRB luminosity relation.
To implement the regression models of ML algorithms, we use scikit-learn\footnote{\url{https://scikit-learn.org/stable/index.html}}, which offers a variety of ML classification and regression models to choose,  e.g.,
Linear Regression(LR), Lasso Regression(Lasso), %Bayesian Ridge(BR), %Gradient Boosting Decision Tree (GBDT), Decision Trees (DT),
Random Forest (RF), Support Vector Regression (SVR), and K-Nearest Neighbors (KNN). %, and Multi-layer Perceptron (MLP), etc. %, and Gaussian Process.
In order to estimate the performance of different ML algorithms that reduce the residuals between real and fitting data,
%we use the values of %the mean absolute error (MAE) and
%the loss function,
%which are given by
we use the values of
the mean squared error (MSE),
which are given by
%\begin{eqnarray}\label{eqnarray6}
%    \mathcal{L}_{\rm SN}= \Delta_{m_B}C_{SN}^{-1}\Delta_{m_B},
%\end{eqnarray}
%\begin{equation}
%% \rm{BVT}=\rm{MSE}-(\frac{1}{\emph{N}}\sum^{N}_{\emph{i}=1}|\emph{Y}_\emph{i}-\hat{\emph{Y}_\emph{i}}|)^2,
%\textrm{MAE} = \frac{1}{N} \sum_{i=1}^{N} |Y_i - \hat{Y}_i|,
%\end{equation}
%and
%\begin{equation}
% \rm{MSE}=\frac{1}{\emph{N}}\sum^{N}_{\emph{\emph{i}}=1}(\emph{Y}_\emph{i}-\hat{\emph{Y}_\emph{i}})^2,
%\textrm{MSE} = \frac{1}{N} \sum_{i=1}^{N} (Y_i - \hat{Y}_i)^2,
%\end{equation}
\begin{equation}
% \rm{MSE}=\frac{1}{\emph{N}}\sum^{N}_{\emph{\emph{i}}=1}(\emph{Y}_\emph{i}-\hat{\emph{Y}_\emph{i}})^2,
\textrm{MSE} = \frac{1}{N} \sum_{i=1}^{N} (Y_i - \hat{Y}_i)^2.
\end{equation}
%\begin{eqnarray}\label{eqnarray6}
%    \mathcal{L}_{\rm SN}= \Delta_{m}C_{SN}^{-1}\Delta_{m},
%\end{eqnarray} where
%\begin{eqnarray}\label{eqnarray6}
%   \Delta_{m}=m^{ML}-m ,
%\end{eqnarray}
%\begin{equation}
%% \rm{MSE}=\frac{1}{\emph{N}}\sum^{N}_{\emph{\emph{i}}=1}(\emph{Y}_\emph{i}-\hat{\emph{Y}_\emph{i}})^2,
%\textrm{L}_{} = \Delta_{m_B}C_{SN}^{-1}\Delta_{m_B} where \Delta_{m_B}=m_B^{ML}-m_B ,
%\end{equation}
%and
%\begin{equation}
%% \rm{BVT}=\rm{MSE}-(\frac{1}{\emph{N}}\sum^{N}_{\emph{i}=1}|\emph{Y}_\emph{i}-\hat{\emph{Y}_\emph{i}}|)^2,
%\textrm{BVT} = \textrm{MSE} - \left(\frac{1}{N} \sum_{i=1}^{N} |Y_i - \hat{Y}_i|\right)^2,
%\end{equation}
%and
%\begin{equation}
% R^2=1-\frac{\sum^{N}_{i=1}(Y_i-\hat{Y_i})^2}{\sum^{N}_{i=1}(Y_i-\overline{Y_i})^2}.
%\end{equation}
%where, $N$ represents the sample size; $Y_i$ represents the $i$-th observation value; $\hat{Y_i}$ is the predicted value for the $i$-th observation. % \textbf{\sout{; $\overline{Y_i}$ represents the average value of observation data}}.
We also select MSE as the evaluation index and utilize the hyperparameter optimization method (GridSearchCV\footnote{\url{https://scikit-learn.org/stable/modules/generated/sklearn.model_selection.GridSearchCV.html}})
provided by scikit-learn to determine optimal hyperparameters for the ML algorithms.
\footnote{In grid search, we assess various hyperparameter combinations for each ML algorithms and utilize the 5-fold cross-validation method to select the one that minimizes MSE as the final configuration}
%The results of MAE and MSE by multiple ML algorithms of scikit-learn are shown in Tab. 1.
%From the analysis results in Tab. 1,

\begin{figure*}
\centering
\includegraphics[width=\textwidth]{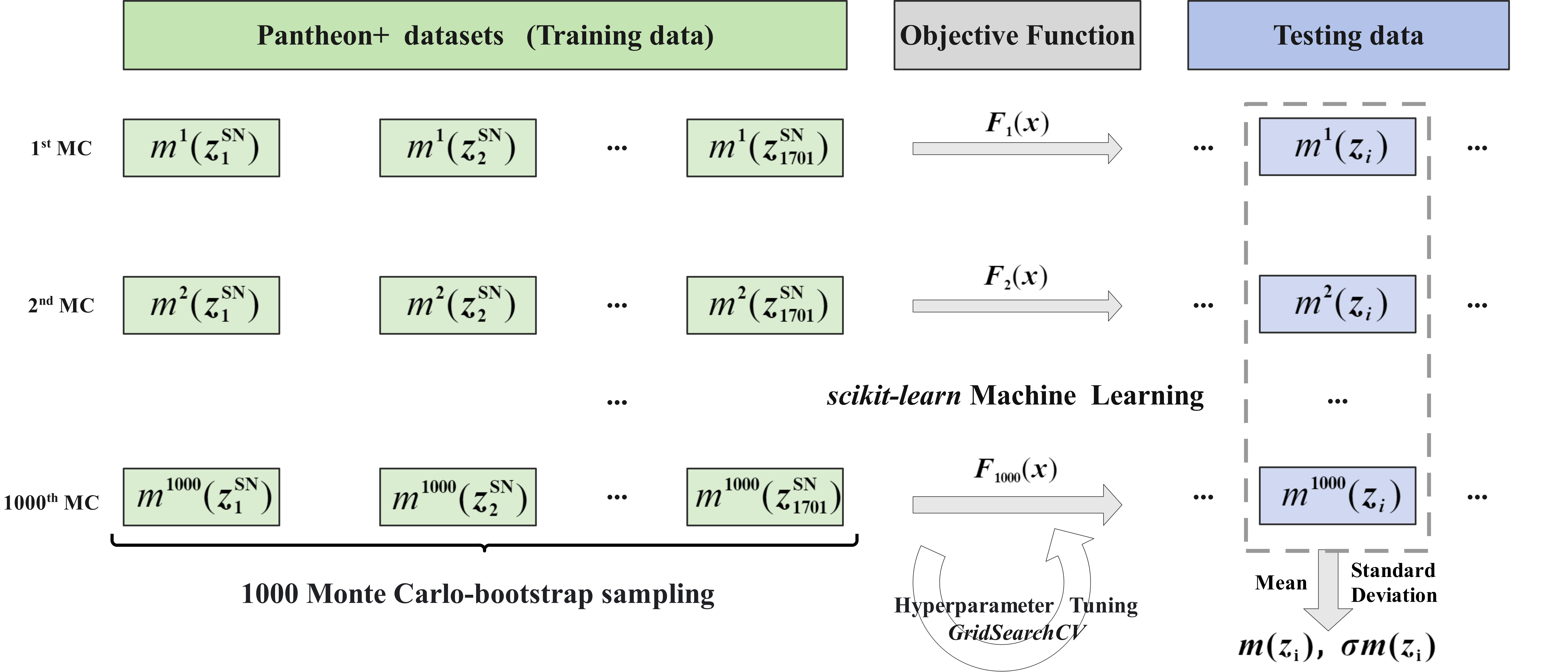}
\caption{The architecture of machine learning algorithms to fit Pantheon+ data.
%These algorithms in scikit-learn do not provide the uncertainties of ML predictions. Therefore,
 To estimate the uncertainty of the apparent magnitude, we use a Monte Carlo-bootstrap method, which sample the apparent magnitude  corresponding to each redshift of SN Ia data  by assuming a normal distribution. %Based on Monte Carlo sampling datasets, 1000 iterations of ML algorithm training were conducted to predict each redshift of the testing set by the objective functions. The average of the prediction results of the objective functions was taken to obtain the $m(z_{i})$ value, and the standard deviation was taken to obtain the error $\sigma_{m(z_{i})}$ value.
%In the training phase, utilizing 1000 Monte Carlo sampled datasets, scikit-learn machine learning algorithms were trained to produce 1000 distinct objective functions. The optimal hyperparameter values for the machine learning algorithm were determined through GridSearchCV methodology; in the testing phase, the trained 1000 objective functions were used to predict each redshift $z_{i}^{test}$ of the test set, and for each redshift $z_{i}^{test}$ point, the average of the prediction results of these 1000 objective functions was taken to obtain the $\mu(z_{i}^{test})$ value, and the standard deviation was taken to obtain the error of the $\sigma_\mu(z_{i}^{test})$ value.
}
\label{model.png}
\end{figure*}

The Pantheon+ dataset consists of 1701 light curves of 1550 unique spectroscopically confirmed SNe Ia ($z$ = 0.00122 to 2.26137), with a table of size 1701x47 (Pantheon+SH0ES.dat\footnote{\url{https://github.com/PantheonPlusSH0ES/DataRelease/tree/main/Pantheon\%2B_Data/4_DISTANCES_AND_COVAR}}).
%The distance modulus is obtained by calculating $\mu = m_{\rm B,corr} - M_{\rm SH0ES}$.
%Following \cite{Perivolaropoulos2023}, we set the absolute magnitude $M_{\rm SH0ES} = -19.235$ \citep{Riess2022} for the SH0ES Cepheid host distance.
It also consists of a 1701x1701 covariance matrix $C_{\rm stat+syst}$  which represents the covariance between SN Ia due to systematic and statistical %distance moluli
uncertainties. %The error of the distance modulus $\mu$ is determined with the covariance matrix between SN Ia due to the uncertainty of the system and statistical distance modulus $C_{\rm stat+syst}$ provided by the Pantheon+ dataset.
%These algorithms in scikit-learn do not provide the uncertainties of ML predictions. Therefore, we use a Monte Carlo-bootstrap method to estimate the uncertainty of the apparent magnitude. Monte Carlo-bootstrap sample the apparent magnitude  corresponding to each redshift in Pantheon+ data assuming a normal distribution. Based on Monte Carlo sampling datasets, 1000 iterations of ML algorithm training were conducted to predict each redshift of the testing set by the objective functions. The average of the prediction results of the objective functions was taken to obtain the $m(z_{i})$ value, and the standard deviation was taken to obtain the error $\sigma_{m(z_{i})}$ value.
Although ML algorithms in scikit-learn are able to predict apparent magnitude of SN Ia at a given redshift, they do not provide their uncertainties. Following \cite{Bengaly2023}, we first develop a Monte Carlo-bootstrap (MC-bootstrap) method to generate 1000 instances of data ($m_{\rm MC}$) with the initial candidate sample being drawn from the distribution of Pantheon+'s apparent magnitude ($m_{\rm obs}$) and covariance matrix $C$.
%The probability density function (PDF) of this candidate can be obtained from
%\begin{eqnarray}\label{mu}
%f(m_{\rm MC}|m_{\rm obs},C)=\frac{1}{(2\pi)^{\frac{k}{2}}|C|^{\frac{1}{2}}}e^{(-\frac{1}{2}\Delta_{m}C^{-1}\Delta_{m})},
%\end{eqnarray} where $   \Delta_{m}=m_{\rm MC}-m_{\rm obs}$,
%$k$ represents the sample size, and $|C|$ is the determinant of the covariance matrix. The PDF  informs the  acceptance probability. If the value of PDF  surpasses a randomly drawn threshold from a uniform distribution, the candidate sample is accepted and incorporated into our sample set. If not, the current sample is retained.
Based on Monte Carlo sampling datasets, 1000 iterations of ML algorithms training are conducted to predict each redshift of the testing set by the objective functions. The average of the prediction results of the objective functions is taken to obtain the $m_{\rm pred}$ value, and the standard deviation is taken to obtain the error $\sigma_{m_{\rm pred}}$ value.
The data flow diagram of ML in our study is shown in Fig. \ref{model.png}.

\begin{figure}
\centering
\includegraphics[width=0.66\textwidth]{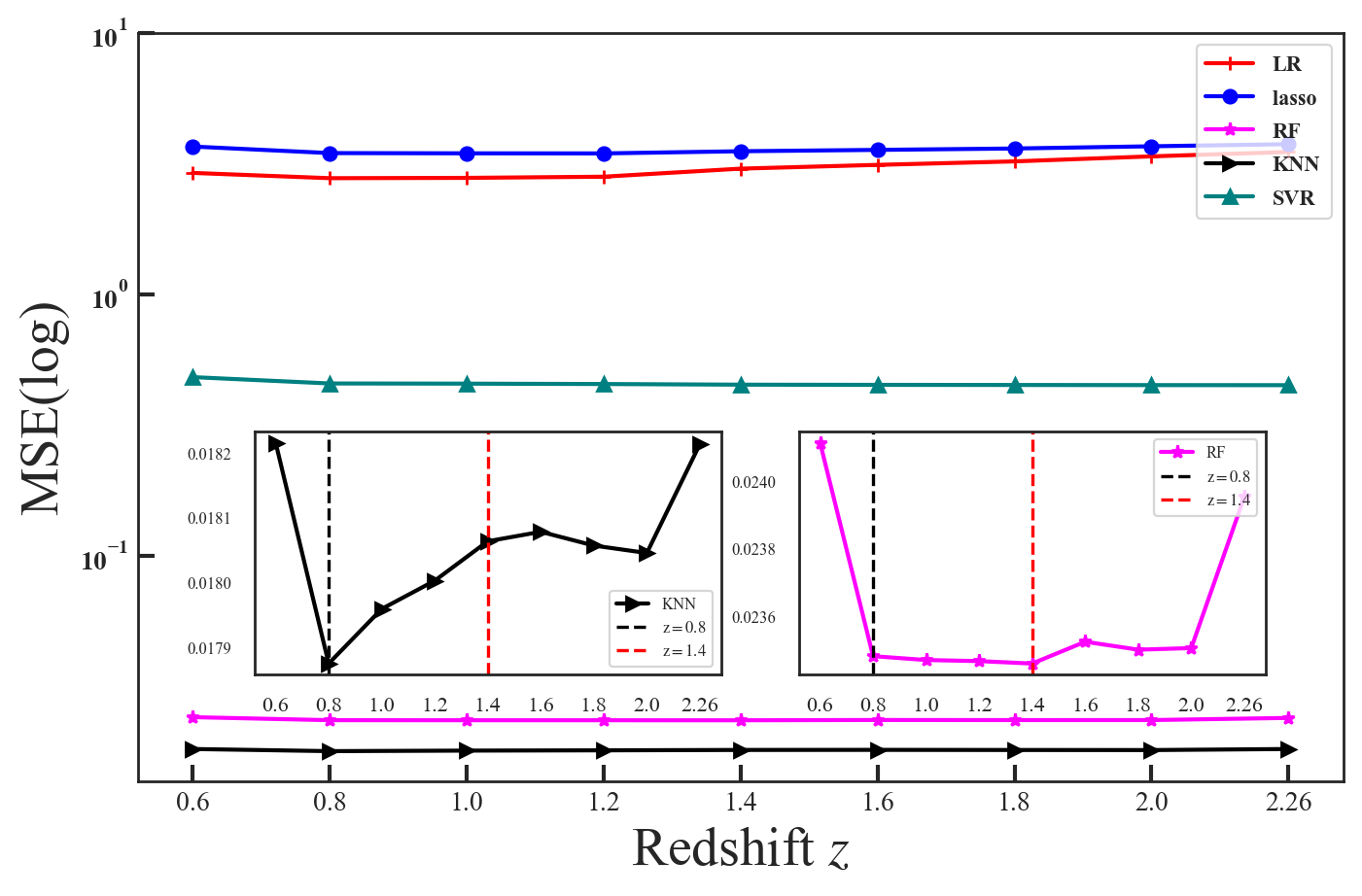}\quad
\caption{The values of MSE  by the multiple ML algorithms fitting Pantheon+ data at redshift splits. Inside plot: Highlight the respective performances of the top two methods with minimal MSE (from left to right: KNN and RF) at redshift splits.} %Enlarge the result of GBDT and KNN in small images}
\label{MSE.fig}
\end{figure}

\begin{figure*}
  \centering
  \includegraphics[width=0.45\textwidth]{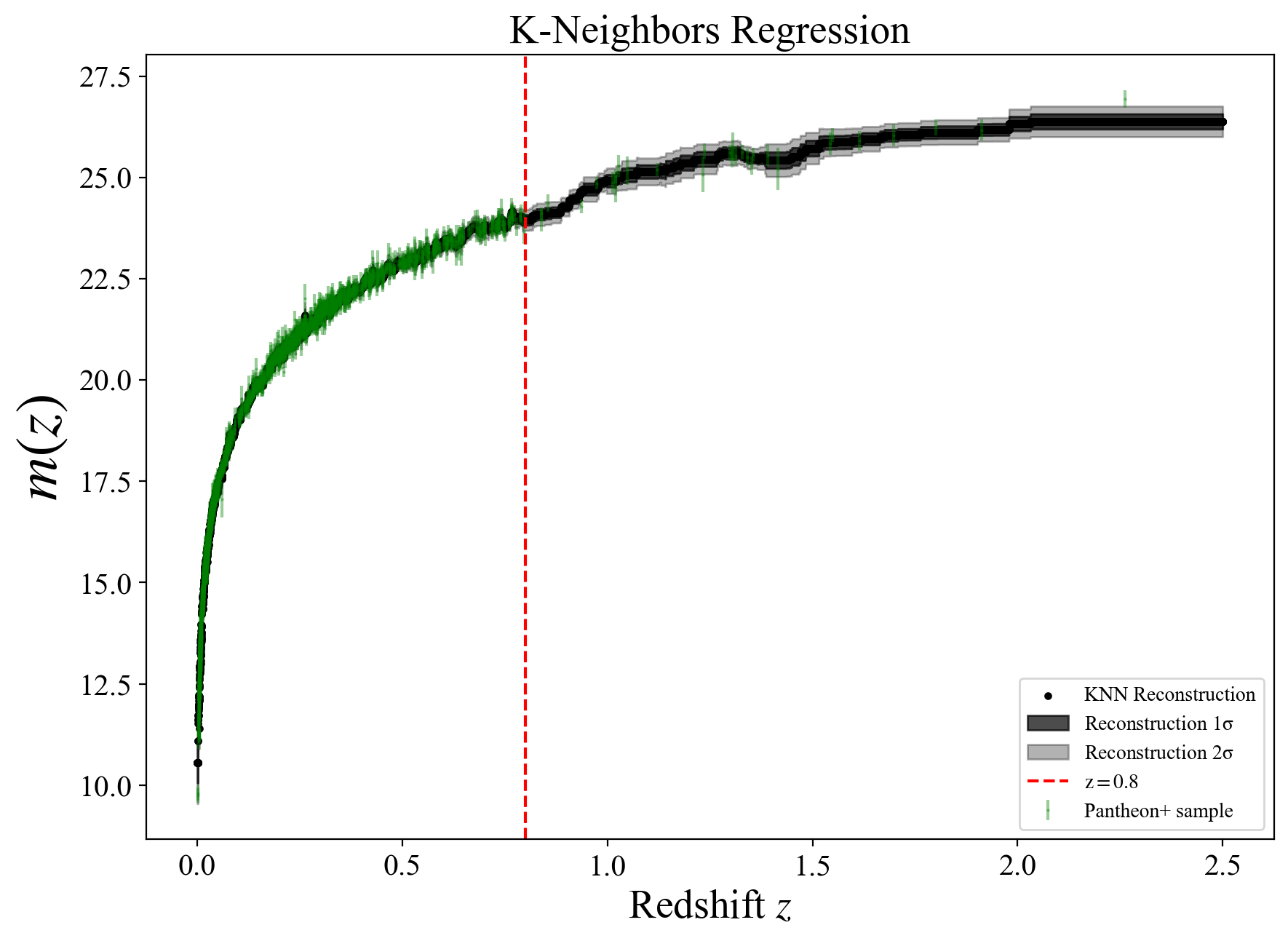} \quad
  \includegraphics[width=0.45\textwidth]{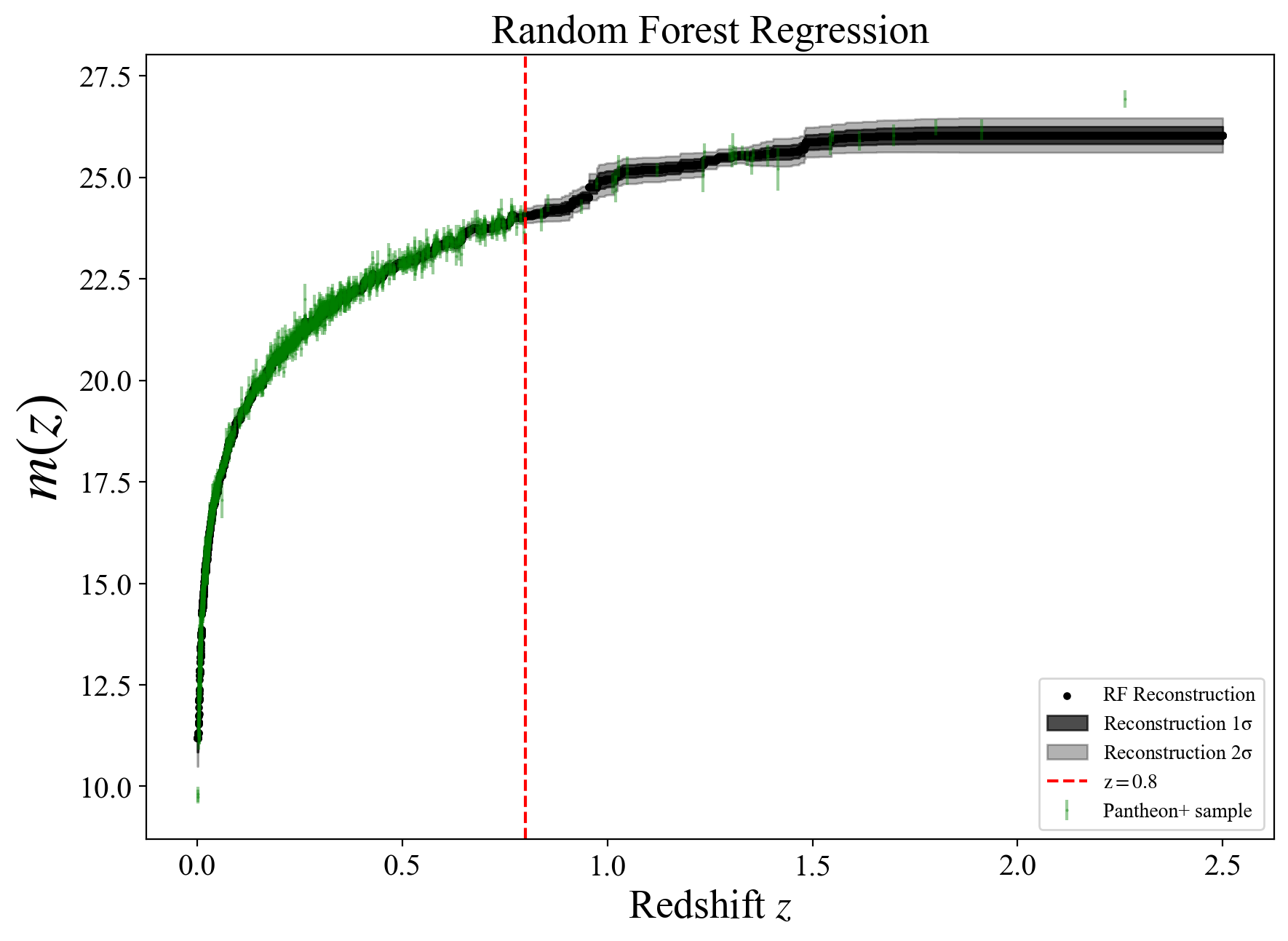} %\quad

  \caption{Reconstruction of the apparent magnitude from the Pantheon+ data set by KNN (left panel) and RF (right panel). The green dots represent Pantheon+ data points with 1$\sigma$ error bars. The model reconstructed center value (black point) and the corresponding 1$\sigma$ and 2$\sigma$ uncertainties (shaded area). %The red curve is the CMB standard distance modulus with $H_0$ = 67.36 km/s/Mpc, $\Omega_{\rm m}$ = 0.315 \citep{Plank2020}, and
  The red dotted line marks the redshift split point corresponding to the minimum MSE at $z=0.8$.}
  \label{recon.png}
\end{figure*}

It should be noted that the Pantheon+ sample do not use SNe from SNLS at $z > 0.8$ due to sensitivity to the $U$ band in model training, therefore the Pantheon+ statistics between $0.8 < z < 1.0$ are lower than that of Pantheon \citep{Scolnic2018} and the Joint Light-curve Analysis (JLA, \cite{Betoule2014}). %we selected different redshift splits (e.g. $z=0.8, 1.4, and 2.26$).
We selected a series of different redshift splits (from 0.6 to 2.26, taking a point every 0.2) %a series of different redshift splits (from 0.6 to 2.0, with intervals $\Delta z= 0.2$)
in the training phase of each ML algorithms to determine the reliable redshift splits of the Pantheon+ sample to calibrate GRBs.
For each redshift split, we take the SNe Ia $z< z_{\rm{split}}$ to calculate  MSE, and the results with the ML algorithms  are shown in Fig. \ref{MSE.fig}.
%Furthermore, as illustrated in the subgraph of Fig. \ref{MSE.fig}, it is observed that the approximate each redshift splits exhibits the smallest MSE when set at 0.8 or 1.4. Therefore,
The results with the ML algorithms at critical redshifts (e.g. $z=0.8, 1.4, 2.26$) are listed in Tab. \ref{Mul ML}.
From Fig. \ref{MSE.fig}, Tab. \ref{Mul ML} (the subgraph of Fig. \ref{MSE.fig}), we find that KNN and RF  methods  show the relatively better performances in redshift splits from 0.6 to 2.26; and all ML methods except SVM achieve the best values at $z=0.8$, which are consistent with the redshift point empirically chosen in \cite{Mu2023} and  \cite{Xie2023} from Pantheon+ sample.
%whereas DT and MLP achieve the best values at $z=1.4$, which is consistent with the redshift point chosen in \cite{Liang2022} from Pantheon sample.
In Fig. \ref{recon.png}, we plot the reconstruction of the apparent magnitude from Pantheon+ by  KNN and RF  methods  with the relatively better performances  reconstructed using the optimal hyperparameters \footnote{When training machine learning algorithms based on each of 1000 Monte Carlo sampled datasets, we employ the GridSearchCV method to determine the optimal hyperparameter values.}. %We also find a significant increase in uncertainty at redshift $z>0.8$. % and SVR model redshift $z >1.4$.

\setlength{\tabcolsep}{0.5mm}{\begin{table*}
\centering
\caption{The corresponding MSE of the multiple ML algorithms from Pantheon+ at critical redshifts $z_{\rm split} = 0.8,1.4,2.26$. \label{Mul ML}}
\begin{tabular}{cccc}
\toprule[1pt]
 ML & \multicolumn{1}{c|}{$z < 0.8 $} & \multicolumn{1}{c|}{$z < 1.4 $} & \multicolumn{1}{c|}{$z < 2.26$} \\
%\cmidrule(lr){2}  \cmidrule(lr){3}  \cmidrule(lr){4}
 ~ & \multicolumn{1}{c|}{(1671~SNe)}  & \multicolumn{1}{c|}{(1693~SNe)}  & \multicolumn{1}{c|}{(1701~SNe)} \\
\midrule[0.8pt]
Lasso  Regression & \multicolumn{1}{c|}{\textbf{3.4771}}  & \multicolumn{1}{c|}{3.5338}  & \multicolumn{1}{c|}{3.7608} \\
\midrule[0.8pt]
%Bayesian Ridge      & \multicolumn{1}{c|}{\textbf{2.7763}}  & \multicolumn{1}{c|}{3.0350} & \multicolumn{1}{c|}{3.5258} \\
%\midrule[0.8pt]
 Linear Regression  & \multicolumn{1}{c|}{\textbf{2.7869}}  & \multicolumn{1}{c|}{3.0312} & \multicolumn{1}{c|}{3.5102} \\
\midrule[0.8pt]
Support Vector Regression & \multicolumn{1}{c|}{0.4565}  & \multicolumn{1}{c|}{0.4516} & \multicolumn{1}{c|}{\textbf{0.4497}}\\
\midrule[0.8pt]
%Multi-layer Perceptron & \multicolumn{1}{c|}{\textbf{0.0499}}  & \multicolumn{1}{c|}{0.0499} & \multicolumn{1}{c|}{0.5003}\\
%\midrule[0.8pt]
Random Forest       & \multicolumn{1}{c|}{\textbf{0.0234}}  & \multicolumn{1}{c|}{0.0234}  & \multicolumn{1}{c|}{0.0239}\\
\midrule[0.8pt]
K-Nearest Neighbors  & \multicolumn{1}{c|}{\textbf{0.0178}} & \multicolumn{1}{c|}{0.0180}  & \multicolumn{1}{c|}{0.0182} \\
\bottomrule[1pt]
\end{tabular}
\end{table*}}

\section{Calibration  of the  Amati relation and the GRB Hubble diagram \label{section3}} %and methodology

%After reconstructing the distance-redshift relation from SNe Ia Pantheon+, we can use it to calibrate the Amati relation.
For GRB data set, we use the A219 sample \citep{Liang2022} with one point GRB051109A removed in the A220 sample \citep{Khadka2021}, which includes the A118 data set with the smallest intrinsic dispersion, as well as 102 data set (A102) from 193 GRBs analyzed by \cite{Amati2019} and \cite{Demianski2017a}. %including the recent Fermi observations.
We divide A219 sample into two subsamples, i.e., the low-redshift GRB sample ($z < 0.8$), which consists of 37 GRBs, and the high-redshift sample ($z > 0.8$), which consists of 182 GRBs.
The Amati relation which connects the spectral peak energy ($E_{\rm p}$) and the isotropic equivalent radiated energy ($E_{\rm iso}$) is expressed as
\begin{equation}y = a + bx \end{equation}
where $y \equiv \log_{10}\frac{E_{\rm iso}}{1{\rm erg}},\quad x \equiv \log_{10}\frac{E_p}{300{\rm keV}}$, %$E_{\rm iso}$ and $E_{\rm p}$ are the isotropic equivalent radiated energy and the spectral peak energy respectively,
$a$ and $b$ are free coefficients; $E_{\rm iso}$ and $E_{\rm p}$ can be respectively expressed as:
\begin{equation}E_{{\rm iso}} = 4\pi d^2_L(z)S_{{\rm bolo}}(1+z)^{-1},\quad E_p = E^{{\rm obs}}_p(1+z) \end{equation}
where $E^{\rm obs}_p$ is the observational value of GRB spectral peak energy and $S_{\rm bolo}$ is observational value of bolometric fluence. %both $E^{{\rm obs}}_p$ and $S_{{\rm bolo}}$ \textbf{are observable}.
The luminosity distance ($d_L$) is related the distance modulus ($\mu$),
$\mu = m - M = 5\log \frac{d_L}{\textrm{Mpc}}+25$.
In order to express the GRB relation direct from the apparent magnitude $m$, %without assuming any prior values of $M$,
we introduce a new coefficient $a'$ to rewrite the Amati relation by
\begin{equation}y' = a'+bx \end{equation}
where $y'=\log_{10}[(1+z)^{-1}({S_{{\rm bolo}}/{1{\rm erg cm^{-2}}}})]+\frac{2}{5}m$, %can be obtained from the observational value of GRB,
$a'=a+2(\frac{M}{5}+5)-\log_{10}[4\pi\textrm{(Mpc/cm)}^{2}]$ and $b$ are free coefficients needing to be calibrated from the GRBs observed data in the formula. Therefore, we can calibrate the Amati relation without assuming any prior values of $M$.

\begin{figure*}
\centering
  \includegraphics[width=0.4\textwidth]{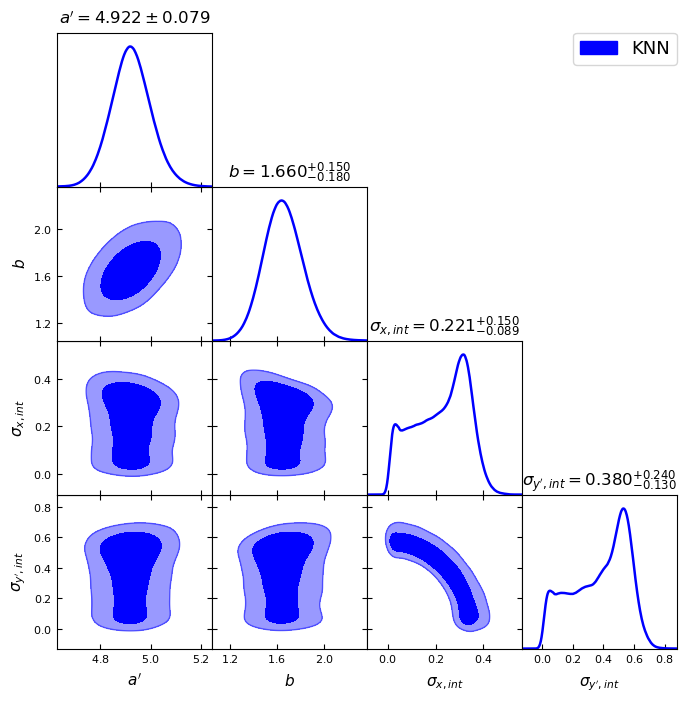} \quad
  \includegraphics[width=0.4\textwidth]{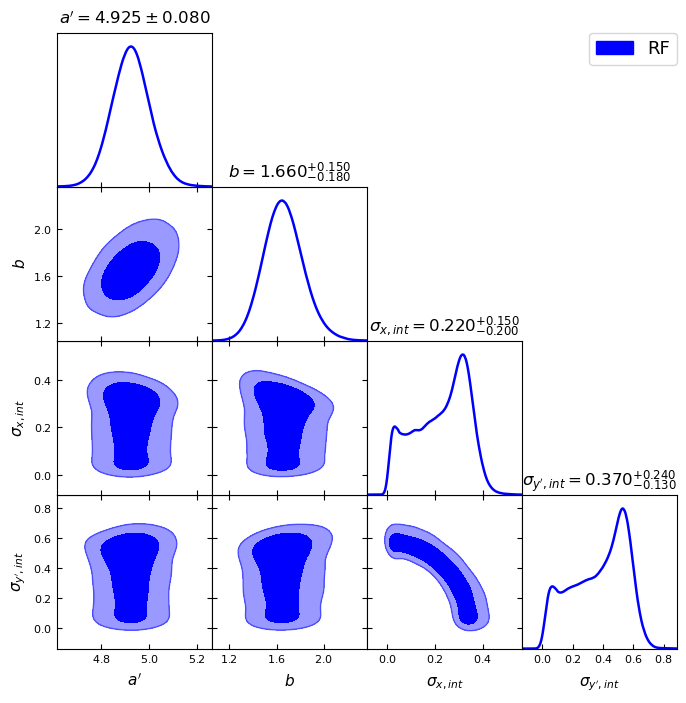}
\caption{The MCMC numerical fitting results (the intercept $a'$, the slope $b$ and the intrinsic scatter along the $x$-axis and $y'$-axis $\sigma_{x,\rm int}$ and $\sigma_{y',\rm int}$ ) of the Amati relation in the A219 GRB sample at $z < 0.8$ (37 GRBs) by the likelihood method \citep{Reichart2001} by KNN (left panel) and RF (right panel).}
\label{Amati.fig}
\end{figure*}

%\begin{figure}
%\centering
%  \includegraphics[width=\hsize,clip]{picture/amtic_ann.png} \quad
% % \includegraphics[width=0.45\textwidth]{picture/amtic_rf.png}
%\caption{The MCMC numerical fitting results (the intercept $a'$, the slope $b$ and the intrinsic scatter along the $x$-axis and $y'$-axis $\sigma_{x,\rm int}$ and $\sigma_{y',\rm int}$ ) of the Amati relation in the A219 GRB sample at $z < 0.8$ (37 GRBs) by the likelihood method \citep{Reichart2001} by \textbf{KNN and ANN}.}
%\label{Amati.fig}
%\end{figure}

We use likelihood function methods \citep{Reichart2001} to fit the parameters of Amati relation
%The likelihood function \citep{D'Agostini2005} is written
%\begin{eqnarray}\label{Lc}
%    \mathcal{L}_{\rm D}\propto\prod_{i=1}^{N_1} \frac{1}{\sigma^2}
%    \times\exp\left[-\frac{[y_i-y(x_i,z_i; a, b)]^2}{2\sigma^2}\right].
%\end{eqnarray}
%Here $\sigma=\sqrt{\sigma_{\rm int}^2+\sigma_{y,i}^2+b^2\sigma_{x,i}^2}$,  $\sigma_{\rm int}$ is the intrinsic scatter of GRBs, $\sigma_x=\frac{1}{\rm ln10}\frac{\sigma_{E_{\rm p}}}{E_{\rm p}},\quad \sigma_y=\frac{1}{\rm ln10}\frac{\sigma_{E_{\rm iso}}}{E_{\rm iso}} $, %$\sigma_{E_{\rm p}}$ is the error magnitude of $E_{\rm p}$,
%and $\sigma_{E_{\rm iso}}=4\pi d^2_L\sigma_{S_{\rm bolo}}(1+z)^{-1}$. % is the error magnitude of $E_{\rm iso}$, where $\sigma_{S_{\rm bolo}}$ is the error magnitude of $S_{\rm bolo}$.
%However, the use of the \cite{D'Agostini2005} likelihood  may introduce a subjective bias on the choice of the independent variable in the analysis.
%The likelihood function of \cite{Reichart2001}
which can be written as \citep{Lin2016,LZL2023}
\begin{eqnarray}\label{eqnarray6}
    \mathcal{L}_{\rm R}\propto\prod_{i=1}^{N_1} \frac{\sqrt{1+b^2}}{\sigma}
    \times\exp\left[-\frac{[y'_i-y'(x_i,z_i; a', b)]^2}{2\sigma^2}\right]
\end{eqnarray}
Here $\sigma=\sqrt{\sigma_{\rm int}^2+\sigma_{y',i}^2+b^2\sigma_{x,i}^2}$,  the intrinsic scatter $\sigma_{\rm int}=\sqrt{\sigma_{y',\rm int}^2 + b^2\sigma_{x,\rm int}^2}$, in which $\sigma_{x,\rm int}$ and $\sigma_{y',\rm int}$ are the intrinsic scatter along the $x$-axis and $y'$-axis.
The likelihood function proposed by \cite{Reichart2001} has the advantage of not requiring the arbitrary choice of an independent variable from $E_{p}$ and $E_{{\rm iso}}$\citep{Amati2013,LZL2023}. \footnote{The use of the \cite{D'Agostini2005} likelihood ($\mathcal{L}_{\rm D}\propto\prod_{i=1}^{N_1} \frac{1}{\sigma^2}
    \times\exp\left[-\frac{[y_i-y(x_i,z_i; a, b)]^2}{2\sigma^2}\right],$
here $\sigma=\sqrt{\sigma_{\rm int}^2+\sigma_{y,i}^2+b^2\sigma_{x,i}^2}$,  $\sigma_{\rm int}$ is the intrinsic scatter of GRBs, $\sigma_x=\frac{1}{\rm ln10}\frac{\sigma_{E_{\rm p}}}{E_{\rm p}},\quad \sigma_y=\frac{1}{\rm ln10}\frac{\sigma_{E_{\rm iso}}}{E_{\rm iso}} $, %$\sigma_{E_{\rm p}}$ is the error magnitude of $E_{\rm p}$,
and $\sigma_{E_{\rm iso}}=4\pi d^2_L\sigma_{S_{\rm bolo}}(1+z)^{-1}$.)  may introduce a subjective bias on the choice of the independent variable in the analysis. The Bivariate Correlated Errors and intrinsic Scatter (BCES) method \citep{AB1996} used in recent Fermi data \citep{WL2024} take into account the possible intrinsic scatter of the data.}

The python package \texttt{emcee} \citep{ForemanMackey2013} %on the basis of the Metropolis-Hastings algorithm
is used to implement the MCMC numerical fitting.
The best fitting parameters $a'$, $b$, $\sigma_{\rm int}$
%\footnote{\textbf{Major comments (6): The results in Figure 4 for GRB baseline trained, show a (high) correlation between the sigmas x and y. This is concerning since wherever statistical analysis develops afterwards, will deteriorate to a high uncertainty (see my comment 2).}}
by  KNN and RF methods
% KNN and RF  methods  with the relatively better performances
%by the ML and GP methods
from the A219 sample at redshift  $z < 0.8$ are shown in Fig. \ref{Amati.fig} and Tab. \ref{Amati result} %\footnote{\textbf{Major comments (4): If the Amati relation is calibrated (Authors should explain with respect to, e.g. H0 or M), this would transfer a model dependence to any baseline in the total $chi^2$-analysis. How then are these results model-independent? Since this is the main conclusion, it is not clear how the Authors got to this conclusion.}}.
For comparison, we also use the \texttt{GaPP} package\footnote{\url{https://github.com/astrobengaly/GaPP}} of the well-known Gaussian process with the squared exponential covariance function \citep{Seikel2012a}.
%\textbf{(Major comments(3): This work should explain in the use of ML (KNN or other) what is the principal  difference with other GP processes reported in the literature.)}
From Tab. \ref{Amati result},
We find that the results of GP are consistent with previous analyses that obtained in \cite{Mu2023} using GaPP from SNe Ia at $z < 0.8$; and the fitting results by GP are consistent with that by  KNN and RF methods in 1$\sigma$ uncertainty, which indicate that ML methods are competitive to GP method.

 \renewcommand{\arraystretch}{1.2} % Ôö¼ÓÐмä¾à
\setlength{\tabcolsep}{1mm}{
\begin{table*}
 \begin{center}{
  \caption{The best-fitting results (the intercept $a'$, the slope $b$ and the intrinsic scatter $\sigma_{\rm int}$ ) of the Amati relation in the A219 GRB sample at $z < 0.8$. %by the likelihood method \citep{Reichart2001}.
  \label{Amati result}}
 \begin{tabular}{cccccc}
 \hline\hline
 %\cmidrule{1-5}
 % Methods & Datasets &$a'$& $b$& $\sigma_{{\rm int}}$ \\
%  \hline
 % KNN & 37GRBs ($z < 0.8$) & $5.070^{+0.094}_{-0.110}$ & $2.29^{+0.22}_{-0.30}$ & $0.649$\\
%  RF & 37GRBs ($z < 0.8$) & $5.069^{+0.093}_{-0.110}$ & $2.30^{+0.22}_{-0.30}$ & $0.643$\\
% % MLP & 37GRBs ($z < 0.8$) & $52.61^{+0.11}_{-0.11}$ & $1.26^{+0.22}_{-0.22}$ & $0.57$\\
%  GaPP  & 37GRBs ($z < 0.8$) & $5.119^{+0.093}_{-0.110}$ & $2.31^{+0.22}_{-0.30}$ & $0.651$ \\
 Likehood & Methods & Datasets &$a'$& $b$& $\sigma_{{\rm int}}$\\
  \hline
  \multirow{3}{*}{\cite{D'Agostini2005}} & KNN & 37GRBs ($z < 0.8$) & $4.830^{+0.074}_{-0.074}$ & $1.26^{+0.15}_{-0.15}$ & $0.551$\\
  ~ & RF & 37GRBs ($z < 0.8$) & $4.831^{+0.074}_{-0.074}$ & $1.26^{+0.15}_{-0.15}$ & $0.551$\\
  ~ & GaPP & 37GRBs ($z < 0.8$)) & $4.880^{+0.11}_{-0.11}$ & $1.26^{+0.22}_{-0.22}$ & $0.574$\\
  \hline
  \multirow{3}{*}{\cite{Reichart2001}} & KNN & 37GRBs ($z < 0.8$) & $4.922^{+0.079}_{-0.079}$ & $1.66^{+0.15}_{-0.18}$ & $0.527$\\
  ~ & RF & 37GRBs ($z < 0.8$) & $4.925^{+0.080}_{-0.080}$ & $1.66^{+0.15}_{-0.18}$ & $0.519$\\
  ~ & GaPP & 37GRBs ($z < 0.8$)) & $4.974^{+0.081}_{-0.081}$ & $1.67^{+0.15}_{-0.18}$ & $0.534$\\
  \hline
  \end{tabular}}
  \end{center}
  \end{table*}}

%From Tab. \ref{Amati result}, we find that the fitting results of the intercept ($a$) with the two likelihood function methods \citep{D'Agostini2005,Reichart2001}  are consistent in 1$\sigma$ uncertainty.
%However, there is a significant difference in the slope parameter ($b$) with the two likelihood function methods, which arises because the likelihood \citep{D'Agostini2005} may introduce subjective biases in the selection of independent variables \citep{LZL2023}.
%and $z \geq 1.4$.

\begin{figure*}
  \centering
  \includegraphics[width=0.45\textwidth]{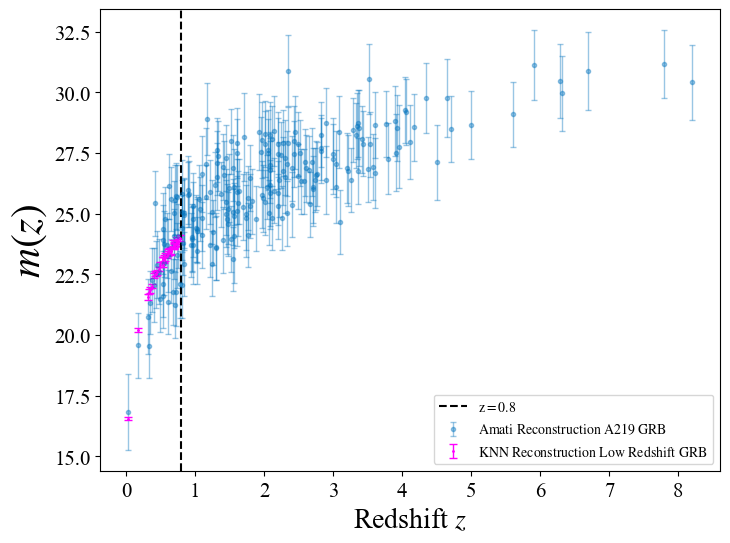} \quad
  \includegraphics[width=0.45\textwidth]{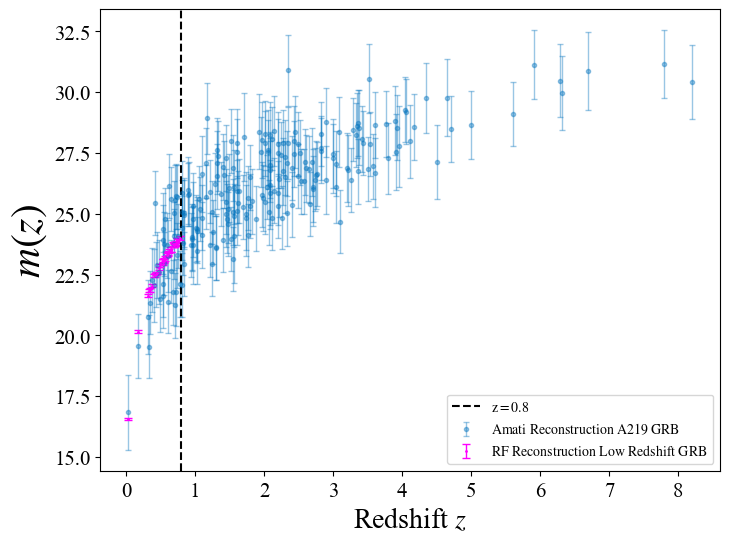} \quad
  \caption{GRB Hubble diagram with the A219 data set. GRBs at $z < 0.8 $ are obtained by the KNN (left panel) and RF (right panel) from the Pantheon+ data, while GRBs with $z \ge 0.8 $ (blue points) are obtained by the Amati relation calibrated with GRBs at $z < 0.8$ by the likelihood method \citep{Reichart2001}. %The red solid  curve is the distance modulus predicted by CMB data based on the $\Lambda$CDM model at  very high-redshift \citep{Plank2020}: $H_0$ = 67.36 km $s^{-1}$ ${{\rm Mpc}}^{-1}$, $\Omega_m$ = 0.315, and the red long dotted curve is the distance modulus fitted by SNe Ia at very low-redshift \citep{Scolnic2022}: $H_0$ = 74.3 km $s^{-1}$ ${{\rm Mpc}}^{-1}$, $\Omega_m$ = 0.298.
  The black dashed line denotes $z = 0.8 $.}
  \label{GRB_Hubble.png}
\end{figure*}

 %Here we assume that the calibration results of the Amati relation at low-redshift are valid at high-redshift; therefore, we can derive the luminosity distances of GRBs at high-redshift to built the GRB Hubble diagram. %The GBRF method uses the likelihood method \citep{D'Agostini2005} to calibrate the Amati relation using A219 $z < 0.8 $ GRBs , and
In order to derive the luminosity distances of GRBs at high-redshift to build the GRB Hubble diagram, we assume that the calibration results of the Amati relation at low-redshift are valid at high-redshift.\footnote{It should be noted that whether the luminosity relations of GRB  are redshift dependent or not is still under debate.
The possible evolutionary effects in GRB relations have been discussed in many works \citep{Lin2016,Wang2017,Demianski2017a,Demianski2021,Dai2021,Tang2021}.
\cite{Khadka2021} found that the Amati relation is independent of redshift within the error bars.
\citep{Liu2022a,Liu2022b} proposed the improved Amati relation by accounting for evolutionary effects via copula, and  %calibrated the copula relations from SNe Ia by the interpolation method to constrain cosmological models, they
found that a redshift evolutionary correlation is slightly favored.
\cite{Jia2022} found no statistically significant evidence for the redshift evolution  with the Amati relation from the analysis of data in different redshift intervals with the 221 GRB sample.
\cite{Kumar2023} calibrated the Amati relation into five redshift bins and find that GRBs seem to evolve with redshift.
Further examinations of possible evolutionary effects should be required for considering GRBs as standard candles for a cosmological probe.}
We utilize the calibration results obtained through the likelihood \citep{Reichart2001} to construct the GRB Hubble diagrams at $z \geq 0.8$ for avoiding any bias in the selection of independent variables.
The Hubble diagram (the apparent magnitude verse the redshift) of A219 GRB sample by  KNN and RF is plotted in Fig. \ref{GRB_Hubble.png}.
%Following \cite{Perivolaropoulos2023}, we set the absolute magnitude $M_{\rm SH0ES} = -19.235$ \citep{Riess2022} for the SH0ES Cepheid host distance.
The uncertainty of the apparent magnitude with the Amati relation can be expressed as %\footnote{\textbf{Major comments (7): The $\mu(z)$ results at high redshift, shows more than $3\sigma$ on the reconstruction throught ML. The analysis is not well described in this issue. While it is possible to obtain the best fit for each point, there is no error propagation (or training) calculated in the full analysis.}}
\begin{eqnarray}\label{eqnarray6}
    \sigma^{2}_{m} = (\frac{5}{2}\sigma_{y'}(a',b,x,\sigma_{{\rm int}}))^{2} + (\frac{5}{2\ln10}\frac{\sigma_{S_{{\rm bolo}}}}{S_{{\rm bolo}}})^{2}
\end{eqnarray}
where
\begin{eqnarray}\label{eqnarray7}
    \sigma^{2}_{y'}(a', b, \sigma_{{\rm int}}, x) = \sigma^{2}_{{\rm int}} + (\frac{b}{\ln10}\frac{\sigma_{E_{p}}}{E_{p}})^{2} + \sigma^{2}_{y'}(a',b);
\end{eqnarray}
Here $\sigma^{2}_{y'}(a',b)=(\frac{\partial{y'}}{\partial a'})^2\sigma^{2}_{a'}+(\frac{\partial{y'}}{\partial b})^2\sigma^{2}_b
+2(\frac{\partial{y'}}{\partial a'})(\frac{\partial{y'}}{\partial b})C^{-1}_{a'b}$, and the inverse of covariance matrix from the fitting coefficients is $(C^{-1})_{a'b}= (\frac{\partial{L}^2}{\partial a'\partial b})$.

\section{Constraints on cosmological models \label{section4}}  %and Discussion
%\subsection{Results from GRBs and OHD}
We use the GRB data in the Hubble diagram at high-redshift to constrain  cosmological models.
The cosmological parameters can be fitted by minimizing the $\chi^2$ statistic.
The $\chi^2$ function for the GRB data can be expressed as %\footnote{\textbf{Major comments (1): This comment is also for the GRB baseline, since in Eq.(8) $\mu$ is already given due a previous calibration.}}
\begin{equation}\chi^2_{\rm GRB} = \sum^{n_1}_{i=1} \left[\frac{m_{\rm obs}(z_i)-m_{\rm th}(z_i;\textbf{\it P},H_0)}{\sigma_{m_i}}\right]^2.
\end{equation}
Here, $n_1$ = 182 is the number of GRBs at $z > 0.8$ in the A219 sample, $m_{{\rm obs}}$ is the observational value of the apparent magnitude with its error \textbf{$\sigma_{m_i}$}, and $m_{{\rm th}}$ is the theoretical value of the apparent magnitude calculated from the cosmological parameters $\textbf{\it P}$, %which can be calculated as
%\begin{eqnarray}\label{mu}
%\mu_{\rm th}=5\log \frac{d_L}{\textrm{Mpc}} + 25=5\log_{10}D_L-\mu_0,
%\end{eqnarray}
%where $\mu_0=5\log_{10}h+42.38$, $h=H_0/(100{\rm km/s/Mpc})$, $H_0$ is the Hubble constant, the unanchored luminosity
%distance $D_L=d_LH_0$.
%\begin{equation}
%\label{chiGRB}
%\chi^2_{\rm GRB}=\Delta \mathbf{m}_{\rm GRB}\mathbf{C}_{\rm GRB}^{-1}\Delta\mathbf{m}_{\rm GRB},
%%\chi^2_{\rm GRB} = \sum^{n_1}_{i=1} \left[\frac{\mu_{\rm obs}(z_i)-\mu_{\rm th}(z_i;\textbf{\it P},H_0)}{\sigma_{\mu_i}}\right]^2.
%\end{equation}
%where $\mathbf{C}_{\rm GRB}$ is the covariance matrix of GRB data, $\Delta m_{\rm GRB}$%\equiv \mu_{\rm GRB}-\mu_{\rm th}(\textbf{\it P})$
%is the vector of residuals between observed distance moduli $m_{\rm GRB}$ and the theoretical value of distance modulus $m_{\rm th}(\textbf{\it P})$, which can be calculated from the cosmological parameters $\textbf{\it P}$,
%Here, $n_1$ = 182 is the number of GRBs at $z > 0.8$ in the A219 sample, $\mu_{{\rm obs}}$ is the observational value of distance modulus with its error $\sigma_{\mu_i}$, and $\mu_{{\rm th}}$ is the theoretical value of distance modulus calculated from the cosmological parameters $\textbf{\it P}$, which can be calculated as
\begin{eqnarray}\label{mu}
m_{\rm th}(\textbf{\it P})=5\log \frac{d_L(\textbf{\it P})}{\rm{Mpc}} + 25+M=5\log_{10}D_L(\textbf{\it P})-\mu_0+M,
\end{eqnarray}
where $\mu_0=5\log_{10}h+42.38$, $h=H_0/(100{\rm km/s/Mpc})$, $H_0$ is the Hubble constant, the unanchored luminosity
distance $D_L(\textbf{\it P})=d_L(\textbf{\it P})H_0$.

We consider three  the dark energy (DE) models for a flat space\footnote{The cosmological models have been usually constrained with flat spatial curvature. It should be noted that recently works constrain nonspatially flat models with GRBs and results are promising \citep{Khadka2021,Cao2022a,Luongo2023}. }, the $\Lambda$CDM model with the Equation of State (EoS) $w=-1$, the $w$CDM model ($w=\rm{const}$), and the CPL model  evolving with redshift with a parametrization EoS ($w=w_0+w_az/(1+z)$). In a flat space,
\begin{eqnarray}
d_{L;{\rm th}}=\frac{{c}(1+z)}{H_{\rm 0}}\int_0^z\frac{dz'}{E(z')}
\end{eqnarray}
here $c$ is the speed of light, $E(z)=[\Omega_{\textrm{M}}(1+z)^3+\Omega_{\textrm{DE}}X(z)]^{1/2}$, and  $X(z)=\exp[3\int_0^z\frac{1+w(z')}{1+z'}dz']$, which is determined by,
\begin{eqnarray}\label{eqnarrayCPL}
X(z)=\begin{cases}
1, & \rm{\Lambda CDM} \\
(1+z)^{3(1+w_0)},& w\rm{CDM} \\
(1+z)^{3(1+w_0+w_a)}e^{-\frac{3w_az}{1+z}}, & \rm{CPL} \\
\end{cases}\end{eqnarray}

\setlength{\tabcolsep}{0.2mm}{
\begin{table*}
% \begin{table}[tbhp]
 \begin{center}{%\scriptsize
  \caption{Joint constraints on parameters of $\Omega_m$, $h$, $w_0$ and $w_a$ for the flat $\Lambda$CDM model, $w$CDM model, and CPL model, obtained by using the KNN, RF, and GaPP methods %, the likelihood method \citep{Reichart2001}
  with 182 GRBs ($ z > 0.8 $) + 32 OHD data. } \label{Joint constrain results}
 \begin{tabular}{cccccccc} \hline
 \cmidrule{1-8} Models & Methods & $\Omega_{m}$ & $h$ & $w_0$  & $w_a$ &  $\Delta \rm AIC$  & $\quad \Delta \rm BIC$ \\ \hline

\multirow{2}{*}{$\Lambda$CDM}
%& KNN & 182 GRBs & $0.36^{+0.11}_{-0.34}$ & $0.54^{+0.08}_{-0.15}$ & - & -\\
%& RF & 182 GRBs & $0.37^{+0.12}_{-0.34}$ & $0.53^{+0.08}_{-0.14}$ & - & -\\
%& GaPP &  182 GRBs & $0.35^{+0.10}_{-0.33}$ & $0.57^{+0.09}_{-0.16}$ & - & -\\
%\cline{2-7}
& KNN & $ 0.329^{+0.046}_{-0.068}$ & $\quad 0.709^{+0.038}_{-0.038}$ & - & -  & - & -\\
& RF & $0.330^{+0.046}_{-0.067}$ & $\quad 0.709^{+0.038}_{-0.038}$ & - & -  & - & -\\
& GaPP & $0.329^{+0.048}_{-0.071}$ & $\quad 0.696^{+0.039}_{-0.039}$ & - & -  & - & -\\
\cmidrule{1-8}
\multirow{3}{*}{$w$CDM }
%& KNN & 182 GRBs & $0.37^{+0.10}_{-0.37}$ & $0.53^{+0.05}_{-0.16}$ & $-1.03^{+0.54}_{-0.54}$ & -\\
%& RF & 182 GRBs & $0.36^{+0.10}_{-0.36}$ & $0.52^{+0.06}_{-0.16}$ & $-1.02^{+0.71}_{-0.87}$ & -\\
%& GaPP & 182 GRBs & $0.36^{+0.11}_{-0.36}$ & $0.55^{+0.05}_{-0.17}$ & $-1.01^{+0.67}_{-0.86}$ & -\\
%\cline{2-7}
& KNN  & $0.301^{+0.080}_{-0.057}$ & $\quad 0.721^{+0.048}_{-0.065}$ & $\quad -1.13^{+0.62}_{-0.38}$ & - & 1.1 & 4.6\\
& RF  & $0.300^{+0.081}_{-0.057}$ & $\quad 0.721^{+0.050}_{-0.064}$ & $\quad -1.12^{+0.60}_{-0.41}$ & -  & 1.1 & 4.4\\
& GaPP  & $0.294^{+0.090}_{-0.060}$ & $\quad 0.706^{+0.050}_{-0.070}$ & $\quad -1.10^{+0.67}_{-0.35}$ & -  & 1.1 & 4.5\\
\cmidrule{1-8}
\multirow{3}{*}{CPL }
%& KNN & 182 GRBs & $0.38^{+0.12}_{-0.34}$ & $0.53^{+0.06}_{-0.16}$ & $-0.97^{+0.57}_{-0.57}$ & $-1.01^{+0.58}_{-0.58}$\\
%& RF & 182 GRBs & $0.38^{+0.11}_{-0.34}$ & $0.54^{+0.05}_{-0.17}$ & $-0.97^{+0.57}_{-0.57}$ & $-1.00^{+0.58}_{-0.58}$\\
%& GaPP &  182 GRBs & $0.37^{+0.12}_{-0.34}$ & $0.56^{+0.06}_{-0.17}$ & $-0.96^{+0.90}_{-0.39}$ & $-1.01^{+0.57}_{-0.57}$\\
%\cline{2-7}
& KNN  & $0.341^{+0.072}_{-0.072}$ & $\quad 0.713^{+0.051}_{-0.051}$ & $\quad -1.03^{+0.60}_{-0.45}$ & $\quad -0.97^{+0.58}_{-0.58}$  & 3.5 & 9.5\\
& RF  & $0.340^{+0.072}_{-0.072}$ & $\quad 0.713^{+0.053}_{-0.069}$ & $\quad -1.04^{+0.59}_{-0.48}$ & $\quad -0.97^{+0.58}_{-0.58}$  & 3.5 & 10.2\\
& GaPP  & $0.338^{+0.074}_{-0.074}$ & $\quad 0.701^{+0.053}_{-0.069}$ & $\quad -1.05^{+0.59}_{-0.50}$ &$\quad -0.95^{+0.57}_{-0.57}$  & 3.4 & 10.2\\
 \cmidrule{1-8}
  \hline
 \end{tabular}}
 \end{center}
 \end{table*}
 }

%\begin{figure}
%\centering
%\includegraphics[width=\hsize,clip]{picture/CDM_RR.png}
%\caption{Constraints on parameters of $\Omega_m$, $h$ for the flat $\Lambda$CDM model by the KNN, RF and GaPP methods with 182 GRBs at $z > 0.8$, The Best-fitting the 1$\sigma$ confidence level results using the KNN method.}
%\end{figure}
%%
%\begin{figure}
%\centering
%\includegraphics[width=\hsize,clip]{picture/WCDM_RR.png}
%\caption{Constraints on parameters of $\Omega_m$, $h$, and $w_0$  for the flat $w$CDM model by the KNN, RF and GaPP methods with 182 GRBs at $z > 0.8$, The Best-fitting the 1$\sigma$ confidence level results using the KNN method.}
%\end{figure}
%
%\begin{figure}
%\centering
%\includegraphics[width=\hsize,clip]{picture/CPL_RR.png}
%\caption{Constraints on parameters of $\Omega_m$, $h$, $w_0$ and $w_a$  for the flat CPL model by the KNN, RF and GaPP methods with 182 GRBs at $z > 0.8$, The Best-fitting the 1$\sigma$ confidence level results using the KNN method.}
%\end{figure}

The OHD  can be obtained from the galactic age differential method \citep{Jimenez2002}, which have advantages to constrain cosmological parameters and distinguish dark energy models.
In our analysis, we also use the latest OHD  \citep{LZL2023} to constrain cosmological models,
including the 31 Hubble parameter measurements at $0.07<z<1.965$ \citep{Stern2010,Moresco2012,Moresco2015,Moresco2016,Zhang2014,Ratsimbazafy2017}, and a new point at $z=0.80$ proposed by \cite{Jiao2023} in a similar approach. In this work, we also use the 31 OHD at $0.07<z<1.965$ and one point at $z=0.75$ from \cite{Jiao2023}.\footnote{It should be noted that  \cite{Borghi2022}  obtained  another new OHD at $z=0.75$. Considering these two measurements are not fully independent and their covariance is not clear, we only use the point \cite{Jiao2023}, which takes advantage of the $~1/\sqrt 2$ fraction of systematic uncertainty.  One could either use the data from \cite{Borghi2022} alternatively with other 31 OHD to investigate cosmology \citep{Cao2022,Muccino2023,Favale2023,Kumar2023}.}
The total OHD contain 32 data, including 15 correlated measurements \citep{Moresco2012,Moresco2015,Moresco2016} with the covariance matrix \citep{Moresco2020}. The $\chi^2$ function for the OHD is, %\footnote{\textbf{Major comments (8): This work uses 31 data points for OHD. However, this sample has issues related to the best fit at low redshift, therefore it is suitable to use the covariance matrix of this sample, instead of the average of the sample. Furthermore, the authors mentioned 31 data points, and afterward, they included 32 points.}}
\begin{equation}\label{eq:chi_Hz}
\chi_\mathrm{OHD}^{2}(\bm{p})=\Delta \hat{H}^{T} \mathbf{C}_{H}^{-1} \Delta \hat{H}+\chi_\mathrm{uncor}^{2} .
\end{equation}
Here the difference vector for the 15 correlated measurements between the observed data ($H_\mathrm{obs}$) and the theoretical values ($H_\mathrm{th}(z;\bm{p})\equiv H_0\sqrt{\Omega_{\textrm{M}}(1+z)^3+\Omega_{\textrm{DE}}X(z)}$ ) is: $\Delta\hat{H}=H_\mathrm{th}(z;\bm{p})-H_{\mathrm{obs}}(z)$; $\mathbf{C}_{H}^{-1}$ is the inverse of the covariance matrix; and the $\chi^2$ function for the 17 uncorrelated measurements is
\begin{equation}\chi_\mathrm{uncor}^{2}=\sum_{i=1}^{17} \left[H_\mathrm{th}(z_i;\bm{p})-H_{\mathrm{obs}}(z_{i})\right]^{2}/\sigma_{H, i}^{2}\end{equation}.
%For the OHD data set, the $\chi^2$ has the form
%\begin{equation}\chi^2_{{\rm {OHD}}} =
%\sum^{n_2}_{i=1} \left [\frac{H_{\rm  obs}(z_i)-H_{\rm th}(z_i;p,H_0)}{\sigma_{H_i}} \right]^2 . \end{equation}
%Here $n_2=32$  denotes  the number of the Hubble parameter measurements. $H_{{\rm obs}}$ is the observational value of Hubble rate and its error $\sigma_{H_i}$. The theoretical Hubble rate of DE models can be calculated by
%\begin{eqnarray}\label{hz}
%H_{\rm{th}}=H_0\sqrt{[\Omega_{\textrm{M}}(1+z)^3+\Omega_{\textrm{DE}}X(z)]}
%\end{eqnarray}

The total $\chi^2$ with the joint data of GRB+OHD can be expressed as
$\chi^2_{{\rm total}} = \chi^2_{{\rm GRB}} +  \chi^2_{{\rm OHD}}.$
The python package \texttt{emcee} \citep{ForemanMackey2013} for the MCMC numerical fitting is used to constrain DE models from the GRB. The cosmological parameters can be fitted by using the minimization $\chi^2$ method through MCMC method. The joint results from 182 GRBs (A219) $z > 0.8 $ with 32 OHD  are shown in Fig. 6 ($\Lambda$CDM), Fig. 7 ($w$CDM) and Fig. 8 (CPL).
%\footnote{\textbf{Minor comments (b): Figures 7-8 show no convergence for the free parameters in standard dark energy parameterisations. The Authors should explain why these models tested with the baselines like SN and GRB are not converging.}}, which are summarized in Tab. 3 \footnote{\textbf{Major comments (2): The fact that the uncertainties on the free dark energy parameters, e.g. w0 and wa, are high, around 10\% on average, is due to the use of A219 GRB at low z. Therefore, the Authors should present a strong reason to use this GRB baseline.}}

\begin{figure}
\centering
\includegraphics[width=0.35\textwidth]{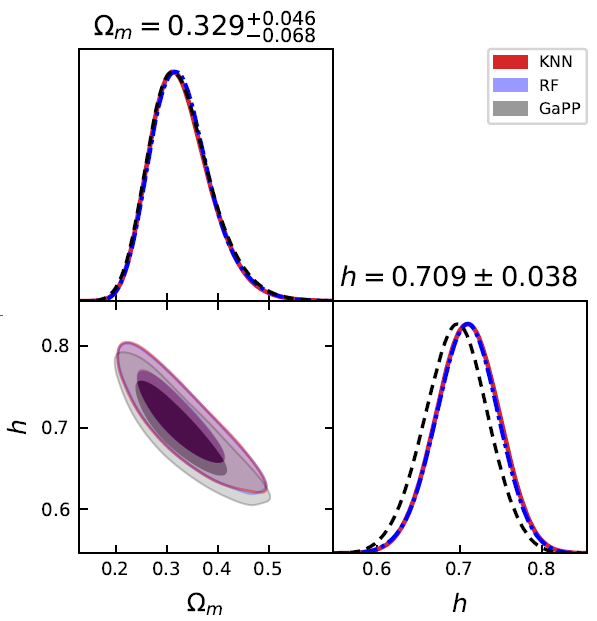}
\caption{Joint constraints on parameters of $\Omega_m$, $h$ for the flat $\Lambda$CDM model by the KNN, RF and GaPP methods with 182 GRBs ($z > 0.8$ ) + 32 OHD. %The best-fitting the 1$\sigma$ confidence level results using the KNN method.
}\label{Hubble_LambdaCDM}
\end{figure}

\begin{figure}
\centering
\includegraphics[width=0.42\textwidth]{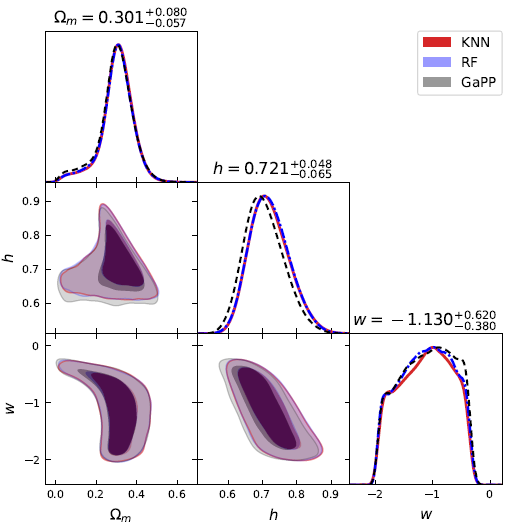}
\caption{Joint constraints on parameters of $\Omega_m$, $h$, and $w_0$  for the flat $w$CDM by the KNN, RF and GaPP methods with 182 GRBs ($z > 0.8$ ) + 32 OHD. %The best-fitting the 1$\sigma$ confidence level results using the KNN method.
} \label{Hubble_wCDM}
\end{figure}

\begin{figure}
\centering
\includegraphics[width=0.5\textwidth]{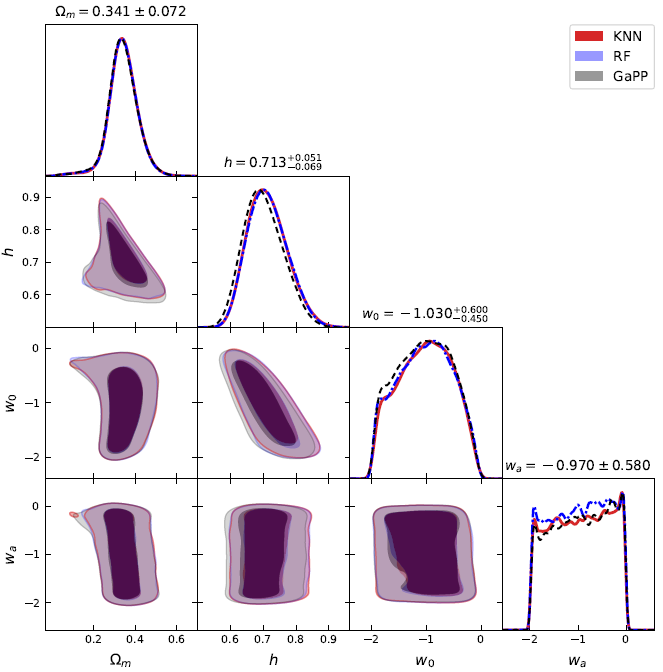}
\caption{Constraints on parameters of $\Omega_m$, $h$, $w_0$ and $w_a$  for the flat CPL model by the KNN, RF and GaPP methods with 182 GRBs ($z > 0.8$ ) + 32 OHD.% The best-fitting the 1$\sigma$ confidence level results using the KNN method.
}
\end{figure}
We find the joint results by the KNN method  are most identical with results by the RF algorithm with 182 GRBs at $0.8<z<8.2$ in the A219 sample and 32 OHD.  By the KNN method, we obtained $\Omega_{\rm m}$ = $0.315^{+0.053}_{-0.076}$ and $h$ = $0.653^{+0.04}_{-0.04}$ for the flat $\Lambda$CDM model;  $\Omega_{\rm m}$ = $0.265^{+0.120}_{-0.067}$, $h$ =  $0.654^{+0.047}_{-0.065}$ , $w$ =  $-1.02^{+0.67}_{-0.31}$ for the flat $w$CDM model; and $\Omega_{\rm m}$ = $0.312^{+0.088}_{-0.068}$ , $h$ =  $0.653^{+0.052}_{-0.073}$  , $w_0$ = $-0.99^{+0.75}_{-0.43}$, $w_a$ = $-0.97^{+0.58}_{-0.58}$ for the CPL model at the 1$\sigma$ confidence level, which favor a possible DE evolution ($w_a\neq0$).
For comparison, we also use the calibration results of \texttt{GaPP} to constrain cosmological models, which are consistent with
the results by KNN and RF with slight difference.
We also find that the results by GP from the Pantheon+ data at $z < 0.8$ are consistent with previous analyses that obtained %in \cite{Mu2023} using GaPP from the Pantheon data+ at $z < 0.8$, as well as %are almost identical %for both \citep{D'Agostini2005} and \citep{Reichart2001} likelihood methods,
in \cite{Liang2022} using GP from the Pantheon data  at $z < 1.4$ for the $\Lambda$CDM model and the $w$CDM model.

For the well-known $H_0$ tension \citep{Hu2023}, $H_0$ with a redshift evolving is an interesting idea\footnote{See e. g.  \citet{Wong2020}, \citet{Krishnan2020}, \citet{Krishnan2021}, \citet{Dainotti2021} for earlier work.}.
\cite{Jia2023} found that $H_0$ value is consistent with that measured from the local data at low redshift  and drops to the value measured from the CMB at high redshift.
Moreover, \cite{Malekjani2023} found the evolving ($H_0$, $\Omega_{\rm m}$) values above $z = 0.7$ in Pantheon+ sample.
Compared to the fitting results from  CMB data based on the $\Lambda$CDM model at  very high-redshift ($H_0$ = 67.36 km $s^{-1}$ ${{\rm Mpc}}^{-1}$, $\Omega_m$ = 0.315)\citep{Plank2020} and SNe Ia at very low-redshift ($H_0$ = 74.3 km $s^{-1}$ ${{\rm Mpc}}^{-1}$, $\Omega_m$ = 0.298)\citep{Scolnic2022}, we find that the $H_0$  value with GRBs by ML at $0.8\le z\le8.2$ and OHD at $z\le1.975$  seems to favor the one from the Planck observations, and the $\Omega_{\rm m}$ value of our results for the flat $\Lambda$CDM model is consistent with the CMB observations at the 1$\sigma$ confidence level.
%We also find a larger $\Omega_{\rm m}$ values in the $\Lambda$CDM model only with GRBs by ML at relative high redshift is obtained, while adding OHD at relative low redshift will remove this trend.

In order to compare the different cosmological models and ML algorithms, we compute the values of the Akaike information criterion (AIC; \cite{Akaike1974,Akaike1981}) and the Bayesian information criterion (BIC; \cite{Schwarz1978}), respectively:
$ \rm AIC = 2\emph{p}-2\ln(\mathcal{L})$,
$ \rm BIC = \emph{p}\ln \emph{N}-2\ln(\mathcal{L})$;
where $\mathcal{L}$ is the maximum value of the likelihood function, $ p $ is the number of free parameters in a model, and $ N $ is the number of data.
% The value of $\Delta \rm AIC$ and $\Delta \rm BIC$, which denotes the difference between $\rm AIC$ and $\rm BIC$  with respect to the reference  model (the $\Lambda$CDM model) are summarized in Table 3.
%For the value of $\Delta \rm AIC$ and $\Delta \rm BIC$,  $0 < \Delta \rm AIC(\Delta \rm BIC) < 2$ indicates difficulty in preferring a given model, $2 < \Delta \rm AIC(\Delta \rm BIC) < 6$ means mild evidence against the given model, and $\Delta \rm AIC(\Delta \rm BIC) > 6$ suggests strong evidence against the model.
We find that the results of $\Delta \rm AIC$ and $\Delta \rm BIC$  by  KNN, RF and GP methods indicate that the $\Lambda$CDM model is favoured respect to the $w$CDM model and the CPL model, which are consistent with the previous analyses \citep{Amati2019} obtained from the 193 GRBs by using the OHD  at $z<1.975$ through the B\'ezier parametric curve combined with 740 SNe Ia. %Meanwhile, we also find that the KNN and RF methods has a lower $-2\ln \mathcal{L}$ value compared to GP, which indicate that ML methods are competitive to GP method.

\section{Results of the Dainotti relation} %and methodology
GRB relations of the prompt emission phase involving the X-ray afterglow plateau phase exist less variability in its features \citep{Dainotti2008, Cardone2009}. In this section,  we also investigate the Dainotti relation\footnote{\cite{Dainotti2008, Dainotti2010, Dainotti2011, Dainotti2013, Dainotti2015, Dainotti2017} proposed the relation between the plateau luminosity and the end time of the plateau in X-ray afterglows (2D Dainotti relation) to constrain cosmological parameters.
Furthermore, \cite{Dainotti2016} proposed the 3D Dainotti relation among the rest-frame time and X-ray luminosity at the end of the plateau emission and the peak prompt luminosity %($T^*_{X}$-$L_{X}$-$L_{\rm{peak}}$)
with small intrinsic scatter.
\cite{Cao2022a,Cao2022b} %used the Amati relation with the A220 and the A118 GRB samples in conjunction with the Dainotti relation to constrain cosmological model parameters by the simultaneous fitting method. They
investigated the 2D and 3D Dainotti relation standardized with the Platinum sample \citep{Dainotti2020} including 50 GRB data. %($L_{X}$-$T^*_{X}$-$L_{\rm{peak}}$),
%and the 3D Dainotti relation is strongly favored over the 2D one with different GRB data compilation.
%$L_0$-$t_b$),
}
by the ML algorithms for comparison.

%\subsection{Calibration  of Dainotti relation}
The Platinum sample \citep{Dainotti2020} listed in Table A1 of \cite{Cao2022a} are used to calibrate the  Dainotti relation %from the Pantheon+ dataset
by the KNN and RF methods. The 2D Dainotti relation which connects the X-ray luminosity $L_{\rm X}$ and the rest-frame time at the end of the plateau emission $T^*_{X}$ is expressed as \citep{Dainotti2008, Dainotti2010, Dainotti2011, Dainotti2013, Dainotti2015, Dainotti2017} and \citep{Cao2022a}
\begin{equation}\log_{10}L_{\rm X} = C_0 + a\log_{10}T^*_{X}\end{equation}
where $C_0$ and $a$ are free coefficients, $L_{\rm X}$ can be calculated by
\begin{equation}L_{\rm X} = {4\pi d^2_{L,\rm rec}(z)F_{{\rm X}}}{(1+z)^{\beta-1}}, \end{equation}
where $F_{{\rm X}}$ is the measured gamma-ray energy flux at $T^*_{X}$, $\beta$ is the X-ray spectral index of the plateau phase in the X-ray band \citep{Evans2009}; $d_{L,\rm rec}$ is related with the reconstructed apparent magnitude %$m_{\rm rec}$
by using the ML algorithms and the absolute magnitude\footnote{Following \cite{Mukherjee2024a, Mukherjee2024b}, we fix the absolute magnitude $M=-19.35$.}.
We use sub-sample at $0.553 \le z < 1.4$  from the Platinum sample, which consists of 50 GRBs ($0.553 \le z \le 5.0$) to calibrate the 2D Dainotti relation.
The results %of the 2D Dainotti relation with the Platinum sample ($z < 1.4 $)
by KNN and RF algorithms are summarized in Table 4. We find that the calibration results by ML from \cite{D'Agostini2005} likelihood function are consistent with those in the current works calibrated with sub-sample at $0.553 \le z \le 1.960$  of the Platinum sample from SNe Ia by neural networks \citep{Mukherjee2024b}:
$a=-1.04\pm0.16, C_0=51.16\pm0.54, \sigma_{\rm int}=0.41\pm0.08$; and from OHD by a Gaussian Processes Bayesian reconstruction tool \citep{Favale2024}: $a=-1.03\pm0.16, C_0=51.20\pm0.52, \sigma_{\rm int}=0.43\pm0.08$.
\renewcommand{\arraystretch}{1.2} % Ôö¼ÓÐмä¾à
\setlength{\tabcolsep}{1mm}{
\begin{table*}
 \begin{center}{
  \caption{The best-fitting results (the slope $a$, the intercept $C_0$  and the intrinsic scatter $\sigma_{\rm int}$ ) of the Dainotti relation  at $z < 1.4 $ by the likelihood method \citep{D'Agostini2005} and \cite{Reichart2001}. }
 \begin{tabular}{cccccc}
 \hline\hline
 %Likehood & ML & data scts &$a$& $b$& $\sigma_{{\rm int}}$ \\
% \hline
%\multirow{3}{*}{\cite{D'Agostini2005}} & KNN  & 9GRBs ($z < 0.8$) & $52.84^{+0.18}_{-0.18}$ & $1.17^{+0.41}_{-0.41}$ & $0.476$ \\
%~ & SVR & 20GRBs ($z < 1.4$) & $52.82^{+0.11}_{-0.11}$ & $1.03^{+0.20}_{-0.20}$ & $0.469$\\
%~ & RF & 9GRBs ($z < 0.8$) & $52.82^{+0.17}_{-0.17}$ & $1.16^{+0.41}_{-0.41}$ & $0.477$\\
%\hline
%\multirow{3}{*}{\cite{Reichart2001}} & KNN & 9GRBs ($z < 0.8$) & $52.91^{+0.12}_{-0.14}$ & $1.91^{+0.30}_{-0.53}$ & $0.483$\\
%~ & SVR & 20GRBs ($z < 1.4$) & $52.80^{+0.082}_{-0.082}$ & $1.42^{+0.14}_{-0.21}$ & $0.454$\\
%~ & RF & 9GRBs ($z < 0.8$) & $52.91^{+0.12}_{-0.14}$ & $1.91^{+0.30}_{-0.53}$ & $0.483$\\
%\hline
 %\cmidrule{1-5}
  Likehood & Methods & Datasets &$a$& $C_0$& $\sigma_{{\rm int}}$ \\
  \hline
  \multirow{2}{*}{\cite{D'Agostini2005}} & KNN & GRBs ($z < 1.4$) & $-1.01^{+0.18}_{-0.18}$ & $51.87^{+0.67}_{-0.67}$ & $0.574$\\
  ~ & RF & GRBs ($z < 1.4$) & $-1.06^{+0.36}_{-0.36}$ & $51.60^{+1.30}_{-1.30}$ & $0.533$\\
  \hline
  \multirow{2}{*}{\cite{Reichart2001}} & KNN & GRBs ($z < 1.4$) & $-1.26^{+0.14}_{-0.13}$ & $52.78^{+0.49}_{-0.49}$ & $0.520$\\
  ~ & RF & GRBs ($z < 1.4$) & $-1.28^{+0.25}_{-0.21}$ & $52.39^{+0.73}_{-0.91}$ & $0.445$\\
 % MLP & 37GRBs ($z < 0.8$) & $52.61^{+0.11}_{-0.11}$ & $1.26^{+0.22}_{-0.22}$ & $0.57$\\
  %GaPP  & 37GRBs ($z < 0.8$) & $5.119^{+0.093}_{-0.110}$ & $2.31^{+0.22}_{-0.30}$ & $0.651$ \\
  \hline
  \end{tabular}}
  \end{center}
  \end{table*}}
%\subsection{Calibration  of Dainotti relation and Constraints on cosmological models}

We combine GRB data at high-redshift ($1.4 \le z \le 5.0 $) with the calibrated 2D Dainotti relation by \cite{Reichart2001} likelihood function to constrain cosmological parameters. The joint results from the high-redshift GRBs and OHD are summarized in Table 5. We find that the results are consistent with analyses that obtained in the calibration with the Amati relation by \cite{Reichart2001} likelihood function in Tab. 3.
\setlength{\tabcolsep}{0.2mm}{
\begin{table*}
% \begin{table}[tbhp]
 \begin{center}{%\scriptsize
  \caption{Joint constraints on parameters of $\Omega_m$, $h$, $w_0$ and $w_a$ for the flat $\Lambda$CDM model, $w$CDM model, and CPL model, obtained by using the KNN, RF %, the likelihood method \citep{Reichart2001}
  with 40 GRBs ($ z > 1.4 $) + 32 OHD data. } \label{Joint constrain results}
 \begin{tabular}{cccccccc} \hline
 \cmidrule{1-8} Models & Methods & $\Omega_{m}$ & $h$ & $w_0$  & $w_a$ &  $\Delta \rm AIC$  & $\quad \Delta \rm BIC$ \\ \hline

\multirow{2}{*}{$\Lambda$CDM}
%& KNN & 182 GRBs & $0.36^{+0.11}_{-0.34}$ & $0.54^{+0.08}_{-0.15}$ & - & -\\
%& RF & 182 GRBs & $0.37^{+0.12}_{-0.34}$ & $0.53^{+0.08}_{-0.14}$ & - & -\\
%& GaPP &  182 GRBs & $0.35^{+0.10}_{-0.33}$ & $0.57^{+0.09}_{-0.16}$ & - & -\\
%\cline{2-7}
& KNN & $ 0.330^{+0.049}_{-0.070}$ & $\quad 0.697^{+0.040}_{-0.040}$ & - & -  & - & -\\
& RF & $0.329^{+0.049}_{-0.071}$ & $\quad 0.694^{+0.041}_{-0.041}$ & - & -  & - & -\\
%& GaPP & $0.316^{+0.053}_{-0.079}$ & $\quad 0.642^{+0.041}_{-0.041}$ & - & -  & - & -\\
\cmidrule{1-8}
\multirow{2}{*}{$w$CDM }
%& KNN & 182 GRBs & $0.37^{+0.10}_{-0.37}$ & $0.53^{+0.05}_{-0.16}$ & $-1.03^{+0.54}_{-0.54}$ & -\\
%& RF & 182 GRBs & $0.36^{+0.10}_{-0.36}$ & $0.52^{+0.06}_{-0.16}$ & $-1.02^{+0.71}_{-0.87}$ & -\\
%& GaPP & 182 GRBs & $0.36^{+0.11}_{-0.36}$ & $0.55^{+0.05}_{-0.17}$ & $-1.01^{+0.67}_{-0.86}$ & -\\
%\cline{2-7}
& KNN  & $0.304^{+0.082}_{-0.060}$ & $\quad 0.709^{+0.052}_{-0.066}$ & $\quad -1.14^{+0.56}_{-0.45}$ & - & 1.7 & 3.9\\
& RF  & $0.299^{+0.084}_{-0.059}$ & $\quad 0.706^{+0.052}_{-0.065}$ & $\quad -1.14^{+0.58}_{-0.44}$ & -  & 1.6 & 3.9\\
%& GaPP  & $0.264^{+0.120}_{-0.069}$ & $\quad 0.646^{+0.048}_{-0.066}$ & $\quad -1.04^{+0.70}_{-0.33}$ & -  & 1.1 & 4.5\\
\cmidrule{1-8}
\multirow{2}{*}{CPL }
%& KNN & 182 GRBs & $0.38^{+0.12}_{-0.34}$ & $0.53^{+0.06}_{-0.16}$ & $-0.97^{+0.57}_{-0.57}$ & $-1.01^{+0.58}_{-0.58}$\\
%& RF & 182 GRBs & $0.38^{+0.11}_{-0.34}$ & $0.54^{+0.05}_{-0.17}$ & $-0.97^{+0.57}_{-0.57}$ & $-1.00^{+0.58}_{-0.58}$\\
%& GaPP &  182 GRBs & $0.37^{+0.12}_{-0.34}$ & $0.56^{+0.06}_{-0.17}$ & $-0.96^{+0.90}_{-0.39}$ & $-1.01^{+0.57}_{-0.57}$\\
%\cline{2-7}
& KNN  & $0.342^{+0.060}_{-0.067}$ & $\quad 0.707^{+0.056}_{-0.068}$ & $\quad -1.09^{+0.49}_{-0.49}$ & $\quad -0.99^{+0.58}_{-0.58}$  & 3.9 & 8.4\\
& RF  & $0.337^{+0.076}_{-0.076}$ & $\quad 0.699^{+0.054}_{-0.067}$ & $\quad -1.06^{+0.60}_{-0.49}$ & $\quad -0.97^{+0.58}_{-0.58}$  & 3.8 & 8.3\\
%& GaPP  & $0.307^{+0.099}_{-0.068}$ & $\quad 0.642^{+0.051}_{-0.072}$ & $\quad -0.98^{+0.74}_{-0.46}$ &$\quad -0.94^{+0.58}_{-0.58}$  & 3.4 & 10.2\\
 \cmidrule{1-8}
  \hline
 \end{tabular}}
 \end{center}
 \end{table*}
 }

\section{Conclusions\label{section5}}

In this paper, we use the ML algorithms to calibrate the Amati relation from the Pantheon+ sample to obtain the GRB Hubble diagram with the A219 sample. The KNN and RF algorithms are selected to calibrate Amati relations due to the best performances.
By the KNN algorithm with GRBs at $0.8<z<8.2$ in the A219 sample and 32 OHD, we obtained $\Omega_{\rm m}$ = $0.315^{+0.053}_{-0.076}$, $h$ = $0.653^{+0.04}_{-0.04}$ for the flat $\Lambda$CDM model;  $\Omega_{\rm m}$ = $0.265^{+0.120}_{-0.067}$, $h$ =  $0.654^{+0.047}_{-0.065}$ , $w$ =  $-1.02^{+0.67}_{-0.31}$ for the flat $w$CDM model; and $\Omega_{\rm m}$ = $0.312^{+0.088}_{-0.068}$ , $h$ =  $0.653^{+0.052}_{-0.073}$  , $w_0$ = $-0.99^{+0.75}_{-0.43}$, $w_a$ = $-0.97^{+0.58}_{-0.58}$ for the CPL model at the 1$\sigma$ confidence level, which are most identical with results by the RF algorithm. These results favor a possible DE evolution ($w_a\neq0$) at the 1-$\sigma$ confidence region for both cases.
We also find that the $\Lambda$CDM model is favoured respect to the $w$CDM model and the CPL model from the results of $\Delta \rm AIC$ and $\Delta \rm BIC$.
Our results with GRBs at $0.8\le z\le8.2$ are consistent with previous analyses that obtained in \cite{Liang2022,Liu2022b,LZL2023} using GP from the Pantheon data and OHD  at $z < 1.4$. %In order to compare with simultaneous fitting method, we also use GRB data sets of A219 sample and SNe Ia to fit the coefficients of the Amati relation ($a$, $b$, $\sigma_{\rm int}$) and the cosmological parameters  ($\Omega_{\rm m}$, $h$, $w$, and $w_a$) simultaneously for the flat $\Lambda$CDM model, the flat $w$CDM model and the flat CPL model. It is found that the simultaneous fitting results are consistent with those obtained from the low-redshift calibration method.
%The reconstruction from cosmological date can be constructed in several ways, e. g., interpolation, polynomials, GP, and so on.
%The main issue in GRB calibration is that we do not know a priori the correct curve to fitting data.
%\cite{Luongo2021} discussed the overall advantage on using machine learning:
%Healing degeneracy and over-fitting issues:
%i) Healing degeneracy and over-fitting issues. More than one model is able to fit the same data will lead to degeneracy in fitting data approaches, and the overall approach of ML overcomes those issues due to interpolation, polynomials with generic over-fitting treatments.
%ii) Speeding up the process of data adaption. ML is a consistency of the data, which automatically encapsulates data without human postulation over their shapes and orders. The complexity of ML models turns out to intimately related to the number of data points. Therefore, the overall process of calibration can be highly improved.
Compared ML to GP, %there are no assumptions of Gaussian distribution for the data. The actual observations may not necessarily conform to Gaussian distributions.
we find that KNN and RF methods with the lowest values in terms of MSE are competitive technics to GP in precision.

Furthermore, we also investigate the Dainotti relation by the ML algorithms for comparison. We find that calibration results of the 2D Dainotti relation are consistent with those in the current works \citep{Mukherjee2024b,Favale2024}; and constrain results at the high-redshift from the  Dainotti relation are consistent with that obtained from the Amati relation.

It should be noted that recent observations from the Dark Energy Spectroscopic Instrument
(DESI) collaboration display slight deviations from $\Lambda$CDM model, see e.g., \cite{Carloni2024,Luongo2024,Colgain2024}.
In future, GRBs could be used to set tighter constraints on cosmological models by the ML technics from recent Fermi data \citep{WL2024} with much smaller scatters, as well as the data from the Chinese-French mission SVOM (the Space-based multiband astronomical Variable Objects Monitor)\citep{Bernardini2021}, which will provide a substantial enhancement of the number of GRBs with measured redshift and spectral parameters.
\section*{ACKNOWLEDGMENTS}
We thank  Zhen Huang, Xin Luo and Prof. Jianchao Feng, Prof. Junjin Peng for kind help and discussions.
%\noindent\textbf{Funding}
This project was supported by the Guizhou Provincail Science and Technology Foundation: QKHJC-ZK[2021] Key 020 and QKHJC-ZK[2024] general 443.
P. Wu was supported by the NSFC under Grants Nos. 12275080, 12073069,
and by the innovative research group of Hunan Province under Grant No. 2024JJ1006, and cultivation project for FAST scientific payoff and research achievement of CAMS-CAS.\\

\noindent\textbf{Data Availability} Data are available at the following references:
the A219 sample of GRB data set
from \cite{Khadka2021,Liang2022},
the Pantheon+ SNe Ia sample from \cite{Scolnic2022},
and the latest OHD obtained with the CC method from \cite{Moresco2020,Moresco2022,Jiao2023} and
\cite{LZL2023}.

\section*{Declarations}

\noindent\textbf{Competing interests} The authors declare no competing interests.\\

\noindent\textbf{Ethics approval} Not applicable.

%\bibliographystyle{plain}
%
%\bibliography{refs}

\begin{thebibliography}{}



\bibitem[Akaike et al.(1974)]{Akaike1974} Akaike, H.
    \href{https://ui.adsabs.harvard.edu/abs/1974ITAC...19..716A/abstract} {1974, ITAC, 19, 716 }
\bibitem[Akaike et al.(1981)]{Akaike1981} Akaike, H.
     \href{https://doi.org/10.1016/0304-4076(81)90071-3} {1981, J. Econ., 16, 3.}
\bibitem[Akritas \& Bershady(1996)]{AB1996} Akritas M.~G.,\& Bershady M.~A.
     \href{https://doi.org/10.1086/177901}{1996, ApJ, 470, 706}
\bibitem[Amati et al.(2019)]{Amati2019} Amati, L., D'Agostino, R., Luongo, O., Muccino, M., \& Tantalo, M.  \href{https://doi.org/10.1093/mnrasl/slz056} {2019, MNRAS, 486, L46}
\bibitem[Amati et al.(2002)]{Amati2002} Amati, L., Frontera, F., Tavani, M., et al. \href{https://doi.org/10.1051/0004-6361:20020722} {2002, A\&A, 390, 81}
\bibitem[Amati et al.(2008)]{Amati2008} Amati, L., Guidorzi, C., Frontera, F., et al. \href{https://doi.org/10.1111/j.1365-2966.2008.13943.x}{2008, MNRAS, 391, 577}
\bibitem[Amati \& Della Valle (2013)]{Amati2013} Amati, L. \& Della Valle
    \href{https://doi.org/10.1142/S0218271813300280}{2013, IJMPD,  22, 1330028}
\bibitem[Arjona(2020)]{Arjona2020} Arjona, R.
    \href{https://doi.org/10.1088/1475-7516/2020/08/009}{2020, JCAP, 08, 009}
\bibitem[Arjona et al.(2021)]{Arjona2021} Arjona, R.,  Lin, H.N.,  Nesseris, S., \& Tang L.
    \href{https://doi.org/10.1103/PhysRevD.103.103513}{2021, PRD,  103, 103513}

\bibitem[Bargiacchi, Dainotti \& Capozziello(2023)]{Bargiacchi2023} Bargiacchi, G., Dainotti, M. G., \&  Capozziello, S.
    \href{https://doi.org/10.1093/mnras/stad2326 }{2023, MNRAS, stad2326}
\bibitem[Betoule et al.(2014)]{Betoule2014} Betoule, M., Kessler, R., Guy, J., et al.
    \href{https://doi.org/10.1051/0004-6361/201423413}{2014, A\&A, 568, A22}
\bibitem[Busti et al.(2014)]{Busti2014}Busti, V. C., Clarkson, C., \& Seikel, M. 2014.
    \href{https://doi.org/10.1093/mnrasl/slu035}{MNRAS, 441, L11}
\bibitem[Benisty(2021)]{Benisty2021} Benisty, D.
    \href{https://doi.org/10.1016/j.dark.2020.100766}{2021, PDU, 31, 100766}
\bibitem[Benisty et al.(2022)]{Benisty2022} Benisty, D., Mifsud, J., Levi Said, J., \& Staicova, D.
    \href{https://arxiv.org/abs/2202.04677}{2022,  arXiv:2202.04677}
\bibitem[Bernardini et al.(2021)]{Bernardini2021} Bernardini, M. G., Cordier, B. \& Wei, J.
    \href{https://doi.org/10.3390/galaxies9040113}{2021, Galaxies, 9, 4}
\bibitem[Bengaly et al.(2023)]{Bengaly2023} Bengaly, C., Dantas , M. A.,  Casarini, L. \& Alcaniz, J.
    \href{https://doi.org/10.1140/epjc/s10052-023-11734-1}{2023, EPJC, 83, 6}

\bibitem[Borghi et al.(2022)]{Borghi2022} Borghi, N., Moresco, M. \& Cimatti, A.
    \href{https://doi.org/10.3847/2041-8213/ac3fb2}{2022, ApJL, 928, L4}
%\bibitem[Breiman(2001)]{Breiman2001} Breiman, L.
%    \href{https://doi.org/10.1023/A:1010933404324}{2001, Machine Learning, 45, 5}
%\bibitem[Breiman(2017)]{Breiman2017} Breiman, L.
%    \href{https://doi.org/10.1201/9781315139470}{2017, Routledge}

\bibitem[Cao et al.(2022a)]{Cao2022a} Cao, S., Dainotti, M., \& Ratra, B.
    \href{https://doi.org/10.1093/mnras/stac517}{2022a, MNRAS, 512, 439}
\bibitem[Cao et al.(2022b)]{Cao2022b} Cao, S., Khadka, N., \& Ratra, B.
    \href{https://doi.org/10.1093/mnras/stab3559}{2022b, MNRAS, 510, 2928}
\bibitem[Cao \& Ratra(2022)]{Cao2022} Cao, S. \& Ratra, B.
    \href{https://doi.org/10.1093/mnras/stac1184}{2022, MNRAS, 513, 5686}
\bibitem[Cao \& Ratra(2024)]{Cao2024} Cao, S. \& Ratra, B.
    \href{https://arxiv.org/abs/2404.08697}{arXiv:2404.08697}
\bibitem[Capozziello \& Izzo(2008)]{Capozziello2008} Capozziello, S., \& Izzo, L.
    \href{https://doi.org/10.1051/0004-6361:200810337}{2008, A\&A, 490, 31}
\bibitem[Capozziello \& Izzo(2009)]{Capozziello2009} Capozziello, S. \& Izzo, L.
    \href{https://doi.org/10.1016/j.nuclphysbps.2009.07.024}{2009, NuPhS, 194, 206}

\bibitem[Capozziello \& Izzo(2010)]{Capozziello2010} Capozziello, S., \& Izzo, L.
    \href{https://doi.org/10.1051/0004-6361/201014522}{2010, A\&A, 519, A73}

\bibitem[Capozziello, D'Agostino, \& Luongo(2018)]{Capozziello2018} Capozziello, S., D'Agostino, R., \& Luongo, O.   \href{https://doi.org/10.1093/mnras/sty422}{2018, MNRAS, 476, 3924}
\bibitem[Cardone et al.(2009)]{Cardone2009} Cardone, V. F., Capozziello, S., \&  Dainotti, M. G.,
    \href{https://doi.org/10.1111/j.1365-2966.2009.15456.x}{2009, MNRAS, 400, 775}
\bibitem[Carloni, Luongo \& Muccino(2024)]{Carloni2024} Carloni, Y., Luongo, O., \& Muccino, M.    \href{}{arXiv:2404.12068}


\bibitem[Chevallier \& Polarski(2001)]{CP2001} Chevallier, M. \& Polarski, D.
    \href{https://doi.org/10.1142/S0218271801000822}{2001, IJMPD, 10, 213}
%\bibitem[Cover \& Hart (1967)]{Cover1967} Cover, T. M., \& Hart, P. E.\href{https://doi.org/10.1109/TIT.1967.1053964}{1967, IEEE Transactions on Information Theory, 13, 21}
\bibitem[Colg\'ain et al.(2024)]{Colgain2024} Colg\'ain, E. O., Dainotti,  M. G.,  Capozziello, S., et al.,
    \href{} {arXiv:2404.08633}
\bibitem[Cucchiara et al.(2011)]{Cucchiara2011} Cucchiara, A., Levan, A., Fox, D. B., et al. \href{https://doi.org/10.1088/0004-637X/736/1/7} {2011, ApJ, 736, 7}
\bibitem[D'Agostini (2005)]{D'Agostini2005} D'Agostini, G.
    \href{https://arxiv.org/abs/physics/0511182v1}{2005,  arXiv: physics/0511182}
\bibitem[Dainotti et al.(2008)]{Dainotti2008} Dainotti, M. G., Cardone V. F., \& Capozziello S. \href{https://doi.org/10.1111/j.1745-3933.2008.00560.x}{2008, MNRAS, L79}
\bibitem[Dainotti et al.(2010)]{Dainotti2010} Dainotti, M. G., Cardone V. F., \& Capozziello S. \href{https://doi.org/10.1088/2041-8205/722/2/L215}{2010, ApJ, 722, L215}
\bibitem[Dainotti et al.(2011)]{Dainotti2011} Dainotti, M. G., Ostrowski, M., \& Willingale, R. \href{https://doi.org/10.1111/j.1365-2966.2011.19433.x}{2011, MNRAS, 418, 2202}
\bibitem[Dainotti et al.(2013)]{Dainotti2013} Dainotti, M. G., Piedipalumbo, E., \& Capozziello S. \href{http://dx.doi.org/10.1093/mnras/stt1516}{2013, MNRAS, 436, 82}
\bibitem[Dainotti et al.(2015)]{Dainotti2015} Dainotti, M. G., Del Vecchio, R., \& Shigehiro, N. \href{https://doi.org/10.1088/0004-637X/800/1/31}{2015, ApJ, 800, 31}
\bibitem[Dainotti et al.(2016)]{Dainotti2016} Dainotti, M. G., Postnikov, S., \& Hernandez, X. \href{https://doi.org/10.3847/2041-8205/825/2/L20}{2016, ApJ, 825, L20}
\textbf{\bibitem[Dainotti et al.(2017)]{Dainotti2017} Dainotti, M. G., Del Vecchio, R., Shigehiro, N., et al.
    \href{}{2017, A\&A, 600, 98}}
\bibitem[Dainotti \& Amati(2018)]{Dainotti2018} Dainotti,M. G., \&  Amati, L.,
    \href{https://doi.org/10.1088/1538-3873/aaa8d7} {2018, PASP, 130, 051001}
\bibitem[Dainotti et al.(2020)]{Dainotti2020} Dainotti, M. G., et al.
    \href{https://doi.org/10.3847/1538-4357/abbe8a}{2020, ApJ, 904, 97}

%\bibitem[Dainotti et al.(2020b)]{Dainotti2020b} Dainotti, M. G., et al.
%    \href{https://doi.org/10.3847/2041-8213/abcda9}{2020b, ApJ, 905, L26}
\bibitem[Dainotti et al.(2021)]{Dainotti2021} Dainotti M.~G., De Simone B., Schiavone T., et al., \href{https://doi.org/10.3847/1538-4357/abeb73}{2021, ApJ, 912, 150}
\bibitem[Dainotti et al.(2022a)]{Dainotti2022a} Dainotti, M. G., Young, S.,  Li, L., et al.
    \href{https://doi.org/10.3847/1538-4365/ac7c64}{2022a, ApJS,  261, 25}
\bibitem[Dainotti et al.(2022b)]{Dainotti2022b} Dainotti, M. G., Nielson, V., Sarracino, G., et al.
    \href{https://doi.org/10.1093/mnras/stac1141}{2022b, MNRAS,  514, 1828 }
\bibitem[Dainotti et al.(2022c)]{Dainotti2022c} Dainotti, M. G., Sarracino G., \&  Capozziello S.
    \href{https://doi.org/10.1093/pasj/psac057}{2022c, PASJ, 74, 1095}
\bibitem[Dainotti et al.(2023)]{Dainotti2023} Dainotti, M. G., Lenart, A. L., Chraya, A., et al.
    \href{https://doi.org/10.1093/mnras/stac2752}{2023, MNRAS, 518, 2201}


\bibitem[Dai et al.(2004)]{Dai2004} Dai, Z., Liang, E., \& Xu, D.
    \href{https://doi.org/10.1086/424694}{2004, ApJ, 612, L101}


\bibitem[Dai et al.(2021)]{Dai2021} Dai, Y., Zheng, X.-G., Li, Z. X., et al. %Gao, H. \& Zhu Z.-H.
    \href{https://doi.org/10.1051/0004-6361/202140895}{2021,  A\&A, 651, L8}
\bibitem[Demianski \& Piedipalumbo(2011)]{DP2011} Demianski, M., \& Piedipalumbo, E.,
    \href{https://doi.org/10.1111/j.1365-2966.2011.18975.x}{2011, MNRAS, 415, 3580}
\bibitem[Demianski et al.(2017a)]{Demianski2017a} Demianski, M., Piedipalumbo, E., Sawant, D., \& Amati, L. \href{https://doi.org/10.1051/0004-6361/201628909}{2017,  A\&A, 598, A112}
\bibitem[Demianski et al.(2017b)]{Demianski2017b} Demianski, M., Piedipalumbo, E., Sawant, D., \& Amati, L. \href{https://doi.org/10.1051/0004-6361/201628911}{2017,  A\&A,598, A113}
\bibitem[Demianski et al.(2021)]{Demianski2021} Demianski, M., Piedipalumbo, E., Sawant, D., \& Amati, L. \href{https://doi.org/10.1093/mnras/stab1669}{2021, MNRAS, 506, 903}
\bibitem[Dinda (2023)]{Dinda2023} Dinda, B. R.
    \href{https://doi.org/10.1142/S0218271823500797}{2023, IJMPD,  32, 2350079}
\bibitem[Dirirsa et al.(2019)]{Dirirsa2019} Dirirsa, F. F., Razzaque, S., Piron, F., et al. \href{https://iopscience.iop.org/article/10.3847/1538-4357/ab4e11/meta}{2019, ApJ, 887, 13}
\bibitem[Dhawan, Alsing \& Vagnozzi(2021)]{Dhawan2021} Dhawan, S., Alsing, J., \& Vagnozzi, S.
    \href{https://doi.org/10.1093/mnrasl/slab058}{2021, MNRAS, 506, L1}



\bibitem[Escamilla-Rivera et al.(2020)]{Escamilla-Rivera2020} Escamilla-Rivera, C., Quintero, M. A. C., \& Capozziello, S. \href{https://doi.org/10.1088/1475-7516/2020/03/008}{2020, JCAP, 03, 008}
\bibitem[Escamilla-Rivera et al.(2022)]{Escamilla-Rivera2022} Escamilla-Rivera, C., Carvajal M., Zamora C., \& Hendry M.
    \href{https://doi.org/10.1088/1475-7516/2022/04/016}{2022, JCAP, 04, 016}
\bibitem[Evans et al.(2009)]{Evans2009} Evans P. A., Beardmore A. P., Page K. L. et al., \href{https://doi.org/10.1111/j.1745-3933.2009.14913.x}{2009, MNRAS , 397, 1177}


\bibitem[Favale et al.(2023)]{Favale2023}Favale, A., Gomez-Valent, A. \& Migliaccio  M.
     \href{https://doi.org/10.48550/arXiv.2301.09591}{2023, arXiv:2301.09591}
\bibitem[Fenimore \& Ramirez-Ruiz(2000)]{Fenimore2000} Fenimore, E. E., \& Ramirez-Ruiz, E. \href{https://arxiv.org/abs/astro-ph/0004176}{2000, arXiv preprint astro-ph/0004176}
\bibitem[Firmani et al.(2005)]{Firmani2005} Firmani, C., Ghisellini, G., Ghirlanda, G., \& Avila-Reese, V. \href{https://doi.org/10.1111/j.1745-3933.2005.00023.x}{2005, MNRAS, 360, L1}
\bibitem[Firmani et al.(2006)]{Firmani2006} Firmani, C., Ghisellini, G., Avila-Reese, V., \& Ghirlanda, G. \href{https://doi.org/10.1111/j.1365-2966.2006.10445.x}{2006, MNRAS, 370, 185}
\bibitem[Favale et al.(2024)]{Favale2024} Favale, A., Dainotti, M. G., Gomez-Valent, A., \& Migliaccio, M. \href{https://doi.org/10.1016/j.jheap.2024.10.010}{2024, JHEA, 44, 323}

\bibitem[Fluri et al.(2018)]{Fluri2018} Fluri, J., Kacprzak, T., Lucchi, A., et al.
    \href{https://doi.org/10.1103/PhysRevD.98.123518}{2018, PRD, 98, 123518}
\bibitem[Fluri et al.(2019)]{Fluri2019} Fluri, J., Kacprzak, T., Lucchi, A., et al.
    \href{https://doi.org/10.1103/PhysRevD.100.063514}{2018, PRD, 100, 063514}
\bibitem[Foreman-Mackey et al.(2013)]{ForemanMackey2013} Foreman-Mackey, D., Hogg, D. W., Lang, D., \& Goodman, J. \href{https://doi.org/10.1086/670067}{2013, PASP, 125, 306}
\bibitem[Gal \& Ghahramani(2016a)]{Gal2016a}Gal, Y., \& Ghahramani, Z.
    \href{https://arxiv.org/pdf/1506.02142.pdf}{2016a, arXiv:1506.02142}
\bibitem[Gal \& Ghahramani(2016b)]{Gal2016b}Gal, Y., \& Ghahramani, Z.
     \href{https://arxiv.org/pdf/1506.02157.pdf}{2016b, arXiv:1506.02157}
\bibitem[Gangopadhyay et al.(2023)]{Gangopadhyay2023} Gangopadhyay, M. R., Sami, M., Sharma, \& Mohit K.
     \href{https://doi.org/10.1103/PhysRevD.108.103526}{2023, PhRvD, 108, 103526}
\bibitem[Gao et al.(2012)]{Gao2012} Gao, H., Liang, N., \& Zhu, Z.-H.
    \href{https://doi.org/10.1142/S0218271812500162}{2012, IJMPD, 21, 1250016}
\bibitem[Ghirlanda et al.(2004a)]{Ghirlanda2004a} Ghirlanda, G., Ghisellini, G., \& Lazzati, D. \href{https://doi.org/10.1086/424913}{2004a, ApJ, 616, 331}
\bibitem[Ghirlanda et al.(2004b)]{Ghirlanda2004b} Ghirlanda, G., Ghisellini, G., Lazzati, D., \& Firmani, C. \href{https://doi.org/10.1086/424915}{2004b, ApJ, 613, L13}
\bibitem[Ghirlanda et al.(2006)]{Ghirlanda2006} Ghirlanda, G., Ghisellini, G.,\& Firmani, C. \href{https://doi.org/10.1088/1367-2630/8/7/123}{2006, New, J. Phys., 8, 123}

\bibitem[G\'{o}mez-Valent \& Amendola(2018)]{Gomez-Valent2018}G\'{o}mez-Valent, A., Amendola, L., 2018.
     \href{https://doi.org/10.1088/1475-7516/2018/04/051}{JCAP, 1804, 051}
\bibitem[Gomez-Valent(2022)]{Gomez-Valent2022} Gomez-Valent, A.
     \href{https://doi.org/}{2022, PhRvD, 105, 043528}
\bibitem[Gowri \& Shantanu(2022)]{Gowri2022} Gowri G. \& Shantanu D.
     \href{https://doi.org/10.1088/1475-7516/2022/10/069}{2022, JCAP, 10, 069}

\bibitem[Hogg et al.(2020)]{Hogg2020} Hogg N.B. , Martinelli M.  \& Nesseris S.
     \href{https://doi.org/10.1088/1475-7516/2020/12/019}{2020, JCAP, 12, 019}
\bibitem[Hu et al.(2021)]{Hu2021} Hu, J. P., Wang, F. Y., \& Dai, Z. G.
     \href{https://doi.org/10.1093/mnras/stab2180}{2021, MNRAS, 507, 730}

\bibitem[Hu \& Wang(2023)]{Hu2023} Hu, J. P. \& Wang, F. Y.
     \href{}{2023, Universe, 9, 94}




\bibitem[Izzo et al.(2015)]{Izzo2015} Izzo, L., Muccino, M., Zaninoni, E., Amati, L., \& Della Valle, M.
     \href{https://doi.org/10.1051/0004-6361/201526461}{2015, A\&A, 582, A115}

\bibitem[Jia et al.(2022)]{Jia2022} Jia, X. D., Hu, J. P., Yang, J., Zhang, B. B., \&  Wang, F. Y.
    \href{https://doi.org/10.1093/mnras/stac2356}{2022, MNRAS, 516, 2575}

\bibitem[Jia et al.(2023)]{Jia2023} Jia, X. D., Hu, J. P. \&  Wang, F. Y.
    \href{}{2023, A\&A, 674, A45}

\bibitem[Jiao et al.(2023)]{Jiao2023} Jiao, K., Borghi, N., Moresco, M. \& Zhang, T-J.
    \href{https://doi.org/10.3847/1538-4365/acbc77}{2023, ApJS, 265, 48}

\bibitem[Jimenez \& Loeb (2002)]{Jimenez2002} Jimenez, R., \& Loeb, A.
    \href{https://doi.org/10.1086/340549}{2002, ApJ, 573, 37}
\bibitem[Jimenez et al.(2003)]{Jimenez2003} Jimenez, R., Verde, L., Treu, T. \& Stern, D. \href{https://doi.org/10.1086/340549}{2003, ApJ, 593, 622}


\bibitem[Kessler \& Scolnic(2017)]{Kessler2017}Kessler, R., \& Scolnic, D.
    \href{https://doi.org/}{2017, ApJ, 836, 56}
\bibitem[Khadka \& Ratra(2020)]{Khadka2020} Khadka, N. \& Ratra, B.
    \href{https://doi.org/10.1093/mnras/staa2779}{2020, MNRAS, 499, 391 }
\bibitem[Khadka et al.(2021)]{Khadka2021} Khadka, N., Luongo, O., Muccino, M., \& Ratra, B. \href{https://doi.org/10.1088/1475-7516/2021/09/042}{2021, JCAP, 09, 042}
\bibitem[Kodama et al.(2008)]{Kodama2008} Kodama, Y., Yonetoku, D., Murakami, T., et al. \href{https://doi.org/10.1111/j.1745-3933.2008.00508.x}{2008, MNRAS, 391, L1}
%\bibitem[Kumar et al.(2022a)]{Kumar2022a} Kumar, D. et al.
%    \href{}{2022, arXiv: 2205.13247}
\bibitem[Krishnan et al.(2020)]{Krishnan2020} Krishnan C., Colg{\'a}in E. {\'O}., Ruchika S., Sheikh-Jabbari M.~M., \& Yang T., \href{https://doi.org/10.1103/PhysRevD.102.103525}{2020, PhRvD, 102, 103525}
\bibitem[Krishnan et al.(2021)]{Krishnan2021} Krishnan C., {\'O} Colg{\'a}in E., Sheikh-Jabbari M.~M., \& Yang T., \href{https://doi.org/10.1103/PhysRevD.103.103509}{2021, PhRvD, 103, 103509}
\bibitem[Kumar et al.(2023)]{Kumar2023} Kumar, D. et al.
    \href{https://doi.org/10.1088/1475-7516/2023/07/021}{2023, JCAP, 07,021}
%\textbf{\bibitem[Lecun et al.(2015)]{Lecun2015} Lecun, Y., Bengio, Y., \& Hinton, G.
%    \href{https://doi.org/10.1038/nature14539}{2015, Nature, 521, 436 }}
\bibitem[Li et al.(2008)]{Li2008} Li, H., Xia, J.-Q., Liu, J., et al.
    \href{https://doi.org/10.1086/529582}{2008, ApJ, 680, 92}
\bibitem[Li \& Lin (2018)]{Li2018}Li, X. and Lin, H.-N., 2018.
    \href{https://doi.org/10.1093/mnras/stx2810}{MNRAS, 474, 313}
\bibitem[Li et al.(2020)]{Li2020} Li, E.-K., Du, M.,  \&  Xu, L.
    \href{https://doi.org/10.1093/mnras/stz3308}{2020, MNRAS, 491, 4960}
\bibitem[Li et al.(2021)]{Li2021} Li, X., Keeley, R. E., Shafieloo, A., et al.
    \href{https://doi.org/10.1093/mnras/stab2154}{ 2021, MNRAS, 507, 919}
\bibitem[Li, Zhang \& Liang (2023)]{LZL2023} Li, Z., Zhang, B.,  \& Liang, N.
    \href{https://doi.org/10.1093/mnras/stad838}{ 2023, MNRAS, 521, 4406}
\bibitem[Li et al.(2023)]{Li2023} Li, J.-L., Yang, Y.-P., Yi, S.-X., Hu, J.-P., Wang, F.-Y., \& Qu, Y.-K.
    \href{https://doi.org/10.3847/1538-4357/ace107}{2023, ApJ, 953, 58}
\bibitem[Liang et al.(2008)]{Liang2008} Liang, N., Xiao, W. K., Liu, Y., \& Zhang, S. N. \href{https://doi.org/10.1086/590903}{2008, ApJ, 685, 354}
\bibitem[Liang \& Zhang (2008)]{LiangZhang2008} Liang, N., \& Zhang, S.
    \href{https://doi.org/10.1063/1.3027949}{2008, AIP Conf. Proc. Vol. 1065, Am. Inst. Phys New York}
\bibitem[Liang et al.(2010)]{Liang2010} Liang, N., Wu, P.,  \& Zhang, S. N.
    \href{https://doi.org/10.1103/PhysRevD.81.083518}{2010, PRD, 81, 083518}
\bibitem[Liang et al.(2011)]{Liang2011}  Liang, N., Xu, L.,  \& Zhu, Z. H.
    \href{https://doi.org/10.1051/0004-6361/201015919}{2011, A\&A, 527, A11}
\bibitem[Liang et al.(2022)]{Liang2022} Liang, N., Li, Z., Xie, X., \& Wu, P.  \href{https://doi.org/10.3847/1538-4357/aca08a}{2022, ApJ, 941, 84}
\bibitem[Liang \& Zhang(2005)]{Liang2005} Liang, E., \& Zhang, B.
    \href{https://doi.org/10.1086/491594}{2005, ApJ, 633, 611}
\bibitem[Liang \& Zhang(2006)]{Liang2006} Liang, E., \& Zhang, B.
    \href{https://doi.org/10.1111/j.1745-3933.2006.00169.x}{2006, MNRAS, 369, L37}
\bibitem[Lin et al.(2016)]{Lin2016} Lin, H. N., Li, X.  \& Chang, Z.
    \href{https://doi.org/10.1093/mnras/stv2471}{2016, MNRAS, 455, 2131 }
\bibitem[Lin et al.(2018)]{Lin2018} Lin, H. N., Li, M. H., \& Li, X.
    \href{https://doi.org/10.1093/mnras/sty2062}{2018, MNRAS, 480, 3117}
\bibitem[Linder(2003)]{Linder2003} Linder, E.V.
    \href{https://doi.org/10.1103/PhysRevLett.90.091301}{2003, PRL,  90, 091301}
\bibitem[Liu et al.(2022a)]{Liu2022a} Liu, Y., Chen, F., Liang, N., et al.
   \href{https://doi.org/10.3847/1538-4357/ac66d3}{2022, ApJ, 931, 50}
\bibitem[Liu et al.(2022b)]{Liu2022b} Liu, Y., Liang, N., Xie, X., et al.
    \href{https://doi.org/10.3847/1538-4357/ac7de5}{2022, ApJ, 935, 7}
\bibitem[Liu \& Wei(2015)]{Liu2015} Liu, J., \& Wei, H.
    \href{https://doi.org/10.1007/s10714-015-1986-1}{2015, GReGr, 47, 141}
\bibitem[Luongo \& Muccino(2020)]{Luongo2020} Luongo, O., \& Muccino, M.
    \href{https://doi.org/10.1051/0004-6361/202038264}{2020, A\&A, 641, A174}
\bibitem[Luongo \& Muccino(2021a)]{Luongo2021a} Luongo, O., \& Muccino, M.
    \href{https://doi.org/10.3390/galaxies9040077}{2021, Galaxies, 9, 77}
\bibitem[Luongo \& Muccino(2021b)]{Luongo2021b} Luongo, O., \& Muccino, M.
    \href{https://doi.org/10.1093/mnras/stab795}{2021, MNRAS, 503, 4581}
%\bibitem[Luongo \& Muccino(2021b)]{Luongo2021b} Luongo, O., \& Muccino, M.    \href{https://doi.org/10.3390/galaxies9040077}{2021, Galaxies, 9, 77}
\bibitem[Luongo \& Muccino(2023)]{Luongo2023} Luongo, O., \& Muccino, M.
    \href{https://doi.org/10.1093/mnras/stac2925}{2023, MNRAS, 518, 2247}

\bibitem[Luongo \& Muccino(2024)]{Luongo2024} Luongo, O., \& Muccino, M.
    \href{}{arXiv:2404.07070}

\bibitem[Malekjani et al.(2023)]{Malekjani2023} Malekjani, M., Mc Conville, R., Colgáin, E. O, et al.
    \href{}{2023,arXiv:2301.12725}


\bibitem[Montiel et al.(2021)]{Montiel2021} Montiel, A., Cabrera, J. I., \&  Hidalgo, J. C.
    \href{https://doi.org/10.1093/mnras/staa3926}{2021, MNRAS, 467, 3239}
\bibitem[Moresco et al.(2012)]{Moresco2012} Moresco, M., Verde, L., Pozzetti, L., Jimenez, R. \& Cimatti, A. \href{https://doi.org/10.1088/1475-7516/2012/08/006}{2012, JCAP, 2012, 053}
\bibitem[Moresco et al.(2015)]{Moresco2015} Moresco, M.
    \href{https://doi.org/10.1093/mnrasl/slv037}{2015, 450, L16}
\bibitem[Moresco et al.(2016)]{Moresco2016} Moresco, M., Pozzetti, L., Cimatti, A. et al.
	\href{https://doi.org/10.1088/1475-7516/2016/05/014}{2016, JCAP, 2016, 014}
\bibitem[Moresco et al.(2020)]{Moresco2020} Moresco M., Jimenez R., Verde L., et al.
    \href{https://doi.org/10.3847/1538-4357/ab9eb0 }{2020, ApJ, 898, 82}
\bibitem[Moresco et al.(2022)]{Moresco2022} Moresco M., Amati, L.,  Amendola, L. et al.
    \href{https://doi.org///10.1007/s41114-022-00040-z}{2022, Living Reviews in Relativity, 25, 6}
\bibitem[Muccino et al.(2021)]{Muccino2021} Muccino, M., Izzo, L., Luongo, O., et al.
    \href{https://doi.org/10.3847/1538-4357/abd254}{2021, ApJ, 908, 181}
\bibitem[Muccino et al.(2023)]{Muccino2023} Muccino, M., Luongo, O., \&  Jain, D.,
    \href{https://doi.org/10.1093/mnras/stad1760} {2023, MNRAS, 523, 4938}
\bibitem[Mu et al.(2023)]{Mu2023} Mu, Y., Chang, B., \& Xu, L.
    \href{https://} {2023, JCAP, 09, 041}
\bibitem[Mukherjee et al.(2024a)]{Mukherjee2024a} Mukherjee, P., Dialektopoulos, K. F., Levi Said, J. \& Mifsud, J.
    \href{https://10.1088/1475-7516/2024/09/060} {2024a, JCAP 09, 060}
\bibitem[Mukherjee et al.(2024b)]{Mukherjee2024b} Mukherjee, P., Dainotti, M. G., \& Dialektopoulos, K. F.
    \href{https://arxiv.org/abs/2411.03773} {2024b, arXiv:2411.03773}


\bibitem[Norris et al.(2000)]{Norris2000} Norris, J. P., Marani, G. F., \& Bonnell, J. T. \href{https://doi.org/10.1086/308725}{2000, ApJ, 534, 248}

\bibitem[Planck Collaboration(2020)]{Plank2020} Planck Collaboration. Aghanim, N., Akrami, Y., Arroja, F., et al. \href{https://doi.org/10.1051/0004-6361/201833880} {2020, A\&A, 641, A1}
\bibitem[Perivolaropoulos \& Skara(2023)]{Perivolaropoulos2023}  Perivolaropoulos, L., \& Skara, F.
    \href{ https://doi.org/10.1093/mnras/stad451} {2023, MNRAS, 520, 5110}

\bibitem[Postnikov et al.(2014)]{Postnikov2014} Postnikov, S., Dainotti, M. G., Hernandez, X., \&  Capozziello, S.
    \href{10.1088/0004-637X/783/2/126}{2014, ApJ, 783, 126}
\bibitem[Shah et al.(2024)]{Shah2024} Shah, R., Saha, S.,  Mukherjee , P. et al.
    \href{ https://doi.org/10.3847/1538-4365/ad5558}{2024, ApJS, 273, 27}
\bibitem[Ratsimbazafy et al.(2017)]{Ratsimbazafy2017} Ratsimbazafy, A. L., Loubser, S. I., Crawford, S. M. et al.
    \href{https://doi.org/10.1093/mnras/stx301}{2017, MNRAS, 467, 3239}


\bibitem[Reichart(2001)]{Reichart2001} Reichart, D. E.
    \href{10.1086/320630}{2001, ApJ, 553, 57}
\bibitem[Riess et al.(2022)]{Riess2022} Riess, A. G., Yuan, W., Macri, L. M., et al.
    \href{https://doi.org/10.3847/2041-8213/ac5c5b}{2022, ApJL, 934, L7}
\bibitem[Rumelhart et al.(1986)]{Rumelhart1986} Rumelhart, D. E., Hinton, G. E., \& Williams, R. J.
    \href{https://doi.org/10.1038/323533a0} {1986, Nature, 323, 533}

\bibitem[Santos-da-Costa et al.(2015)]{Santos-da-Costa2015}Santos-da-Costa, S., Busti, V. C., and Holanda, R. F. L., 2015.
    \href{https://doi.org/10.1088/1475-7516/2015/10/061}{JCAP, 10, 061}
\bibitem[Schaefer(2003)]{Schaefer2003} Schaefer, B. E.
    \href{https://doi.org/10.1086/368104}{2003, ApJ, 583, L67}
\bibitem[Schaefer(2007)]{Schaefer2007} Schaefer, B. E.
    \href{https://doi.org/10.1086/511742}{2007, ApJ, 583, L67}
\bibitem[Scolnic et al.(2018)]{Scolnic2018} Scolnic, D. M., Jones, D. O., Rest, A., et al.
    \href{https://doi.org/10.3847/1538-4357/aab9bb}{2018, ApJ, 859, 101}
\bibitem[Scolnic et al.(2022)]{Scolnic2022} Scolnic, et al.
    \href{https://doi.org/10.3847/1538-4357/ac8b7a}{ 2022, ApJ, 938, 113}
\bibitem[Schwarz et al.(1978)]{Schwarz1978} Schwarz, G.
    \href{https://ui.adsabs.harvard.edu/abs/1978AnSta...6..461S/abstract}{1978, AnSta, 6, 461}
\bibitem[Seikel et al.(2012a)]{Seikel2012a} Seikel, M., Clarkson, C., \& Smith, M.
    \href{https://doi.org/10.1088/1475-7516/2012/06/036}{2012, JCAP, 06, 036}
\bibitem[Seikel et al.(2012b)]{Seikel2012b} Seikel, M., Yahya, S., Maartens, R., \&  Clarkson, C.
    \href{https://doi.org/10.1103/PhysRevD.86.083001}{2012, PRD, 86, 083001}
\bibitem[Seikel \& Clarkson (2013)]{Seikel2013}Seikel, M., \& Clarkson, C., 2013,
    \href{https://arxiv.org/abs/1311.6678}{arXiv:1311.6678}

\bibitem[Shirokov et al.(2020)]{Shirokov2020} Shirokov, S. I., Sokolov, I. V., Lovyagin, N. Yu, et al.
    \href{https://doi.org/10.1093/mnras/staa1548} {2020, MNRAS, 496, 1530}
\bibitem[Simon et al.(2005)]{Simon2005} Simon, J.,Verde, L. \& Jimenez, R.
    \href{https://doi.org/10.1103/PhysRevD.71.123001}{2005, PhRvD, 71, 123001}
%\bibitem[Smola \& Scholkopf(2004)]{Smola2004} Smola, A. J., \& Scholkopf, B.
%    \href{https://doi.org/10.1023/B:STCO.0000035301.49549.88}{2004, Statistics and Computing, 14, 199}
\bibitem[Srivastava et al.(2014)]{Srivastava2014} Srivastava, N., Hinton, G., Krizhevsky, A., Sutskever, I., et al.
    \href{http://jmlr.org/papers/v15/srivastava14a.html}{2014, The journal of machine learning research, 15, 1929}
\bibitem[Stern et al.(2010)]{Stern2010} Stern, D., Jimenez, R., Verde, L., Kamionkowski, M. \& Starford, S. A.
    \href{https://doi.org/10.1088/1475-7516/2010/02/008}{2010, JCAP, 2010, 008}

\bibitem[Tang et al.(2021)]{Tang2021} Tang L., Li X., Lin, H.-N., \& Liu L.
    \href{https://doi.org/10.3847/1538-4357/abcd92} {2021, ApJ, 907, 121}
\bibitem[Tang et al.(2022)]{Tang2022} Tang L., Lin, H.-N., Li X., \& Liu L.
    \href{https://doi.org/10.1093/mnras/stab2932} {2022, MNRAS, 509, 1194}
\bibitem[Tsutsui et al.(2009a)]{Tsutsui2009a} Tsutsui, R., Nakamura, T., Yonetoku, D.,  Murakami, T., Kodama, Y., \& Takahashi, K.    \href{https://doi.org/10.1088/1475-7516/2009/08/015}{2009, JCAP, 0908, 015}

\bibitem[Tsutsui et al.(2009b)]{Tsutsui2009b} Tsutsui, R., Nakamura, T., Yonetoku, D., et al. \href{https://doi.org/10.1111/j.1745-3933.2008.00604.x}{2009, MNRAS, 394, L31}
\bibitem[Vagnozzi, Loeb \& Moresco(2021)]{Vagnozzi2021} Vagnozzi, S., Loeb, A., \& Moresco, M.
    \href{https://doi.org/10.3847/1538-4357/abd4df}{2021, ApJ, 908, 84}

\bibitem[Wang \& Dai(2006)]{Wang2006} Wang, F., \& Dai, Z. G.
    \href{https://doi.org/10.1111/j.1365-2966.2006.10108.x}{2006, MNRAS, 368, 371}
\bibitem[Wang(2008)]{Wang2008} Wang, Y.
    \href{https://doi.org/10.1103/PhysRevD.78.123532}{2008, PhRvD, 78, 123532 }
\bibitem[Wang \& Dai(2011)]{WangDai2011}  Wang, F. Y., \& Dai, Z. G.
    \href{https://doi.org/10.1051/0004-6361/201117517}{2011, A\&A, 536, 96}
\bibitem[Wang et al.(2016)]{Wang2016}  Wang, J. S., Wang, F. Y., Cheng, K. S., \& Dai, Z. G. \href{https://doi.org/10.1051/0004-6361/201526485}{2016, A\&A, 585, A68}
\bibitem[Wang \& Wang(2019)]{Wang2019}  Wang, Y. Y., \& Wang, F. Y.
    \href{https://doi.org/10.3847/1538-4357/ab037b}{2019, ApJ, 873, 39}
\bibitem[Wang et al.(2022)]{Wang2022}  Wang, F. Y., Hu, J. P., Zhang, G. Q., \& Dai, Z. G. \href{https://doi.org/10.3847/1538-4357/ac3755}{2022, ApJ, 924, 97}
\bibitem[Wang et al.(2017)]{Wang2017}Wang, G.-J., Wei, J.-J., Li, Z.-X., et al. 2017.
    \href{https://doi.org/110.3847/1538-4357/aa8725}{ApJ, 847, 45}
\bibitem[Wang et al.(2020)]{Wang2020}Wang, G.-J., Ma, X.-J., Li, S.-Y., \& Xia, J.-Q. 2020.
    \href{https://doi.org/10.3847/1538-4365/ab620b}{ApJS, 246, 13}
\bibitem[Wang et al.(2021)]{Wang2021}Wang, G.-J., Ma, X.-J. \& Xia, J.-Q. 2021.
    \href{https://doi.org/10.1093/mnras/staa4044}{MNRAS, 501, 5714}
\bibitem[Wang, Li \& Liang(2024)]{WLL2024} Wang, G., Li, X., Liang, N.
    \href{https://doi.org/10.1007/s10509-024-04340-4}{2024, ApSS, 369, 74}
\bibitem[Wang \& Liang (2024)]{WL2024} Wang, H. \& Liang, N.
    \href{} {arXiv:2405.14357}
\bibitem[Wei \& Zhang(2009)]{Wei2009} Wei, H., Zhang, S. N.
    \href{https://doi.org/10.1140/epjc/s10052-009-1086-z}{2009, EPJC, 63, 139}
\bibitem[Wei(2010)]{Wei2010} Wei, H.,
    \href{https://doi.org/10.1088/1475-7516/2010/08/020}{2010, JCAP, 08, 020}
\bibitem[Wei \& Wu (2017)]{Wei2017}Wei, J.-J. \& Wu, X.-F., 2017.
    \href{https://doi.org/10.3847/1538-4357/aa674b}{ApJ, 838, 160w}

\bibitem[Wong et al.(2020)]{Wong2020} Wong K.~C., Suyu S.~H., Chen G.~C.-F., Rusu C.~E., et al., \href{https://doi.org/10.1093/mnras/stz3094}{2020, MNRAS, 498, 1420}
\bibitem[Xie et al.(2023)]{Xie2023} Xie, H., Nong, X., Wang, H., Zhang, B., Li, Z. \& Liang, N.
    \href{} {arXiv:2307.16467}
\bibitem[Xu et al.(2005)]{Xu2005} Xu, D., Dai, Z., \& Liang, E.
    \href{https://doi.org/10.1086/466509}{2005, ApJ, 633, 603}

\bibitem[Xu et al.(2021)]{Xu2021} Xu, F., Tang, C.-H., Geng, J.-J., Wang, F.-Y., Wang, Y.-Y., Kuerban, A. \& Huang, Y.-F. \href{https://doi.org/10.3847/1538-4357/ac158a}{2021, ApJ, 920, 135}
\bibitem[Xu et al.(2022)]{Xu2022} Xu, B., Wang, Z., Zhang, K., Huang, Q. \& Zhang, J. \href{https://doi.org/10.3847/1538-4357/ac9793}{2022, ApJ, 939, 115}
\bibitem[Xu et al.(2023)]{Xu2023} Xu, F., Huang, Y.-F. , Geng, J.-J., et al.
    \href{https://doi.org/10.1051/0004-6361/202245414}{2023, A\&A, 673, A20}
\bibitem[Yonetoku et al.(2004)]{Yonetoku2004} Yonetoku, D., Murakami, T., Nakamura, T., et al.
    \href{https://doi.org/10.1086/421285}{2004, ApJ, 609, 935}
\bibitem[Yu, Qi \& Lu(2009)]{Yu2009} Yu, B., Qi, S., \& Lu, T.
    \href{https://doi.org/10.1088/0004-637X/705/1/L15}{2009, ApJ, 705, L15}
\bibitem[Yahya et al.(2014)]{Yahya2014}Yahya, S., Seikel, M., Clarkson, C., Maartens, R., \& Smith, M. 2014.
    \href{https://doi.org/10.1103/PhysRevD.89.023503}{PRD, 89, 023503}
\bibitem[Yang et al.(2015)]{Yang2015}Yang, T, Guo, Z-K, Cai, R-G, 2015.
    \href{https://doi.org/10.1103/PhysRevD.91.123533}{PRD, 91, 123533}
 \bibitem[Yang et al.(2019)]{Yang2019}Yang, T., Holanda, R. F. L., and Hu, B., 2019.
    \href{https://doi.org/10.1016/j.astropartphys.2019.01.005}{Astropart. Phys. 108, 57-62}

\bibitem[Zhang (2014)]{ZhangYi2014}Zhang, Y., 2014.
    \href{https://doi.org/10.48550/arXiv.1408.3897}{arXiv: 1408.3897}
\bibitem[Zhang et al.(2014)]{Zhang2014} Zhang, C., Zhang, H., Yuan, S., Liu, S., Zhang, T. \& Sun, Y.
    \href{https://doi.org/10.1088/1674-4527/14/10/002}{2014, RAA, 14, 1221}
\bibitem[Zhang et al.(2022)]{Zhang2022} Zhang, J. C., Jiao, K., Zhang, T., Zhang, T. J., \& Yu, B.
    \href{https://doi.org/10.3847/1538-4357/ac85aa}{2022, ApJ, 936, 21}

\bibitem[Zhou \& Li (2019)]{Zhou2019}Zhou, H., \& Li, Z., 2019.
    \href{https://doi.org/10.1088/1674-1137/43/3/035103}{Chinese Physics C., 43, 035103}

\end{thebibliography}

\end{document}